\documentclass[12pt]{article}
\usepackage{mathtools,amssymb,mathrsfs,microtype,tikz,ytableau}
\usepackage[colorlinks=true]{hyperref}
\pdfoutput=1

\usepackage[backend=bibtex,style=numeric-comp,sorting=none,maxbibnames=99, minbibnames=99]{biblatex}
\ytableausetup{centertableaux,boxsize=.5em}

\usepackage{slashed}

\makeatletter\@addtoreset{equation}{section}\makeatother

\DeclareMathOperator{\Tr}{tr}
\DeclareMathOperator{\sign}{sign}
\DeclareMathOperator{\re}{re}
\DeclareMathOperator{\im}{im}

\renewcommand{\title}[1]{\vbox{\center\LARGE{#1}}\vspace{5mm}}
\renewcommand{\author}[1]{\vbox{\center\large#1}\vspace{5mm}}
\newcommand{\address}[1]{\vbox{\center\em#1}}

\begin{document}
 
\begin{titlepage}
\begin{center}
\hfill {\tt }\\
\vspace{1mm}

\title{
 {\LARGE Dynamics of QCD$_{3}$ with Rank-Two Quarks And Duality}}
\vspace{4mm}

Changha Choi,${}^{a}$\footnote{\href{mailto: changha.choi@stonybrook.edu}
{\tt changha.choi@stonybrook.edu}}
Diego Delmastro,${}^{bc}$\footnote{\href{mailto:ddelmastro@perimeterinstitute.ca}
{\tt ddelmastro@perimeterinstitute.ca}}
Jaume Gomis,${}^{b}$\footnote{\href{mailto:jgomis@perimeterinstitute.ca}
{\tt jgomis@perimeterinstitute.ca}}
Zohar Komargodski,${}^{de}$\footnote{\href{mailto:zkomargo@gmail.com }
{\tt zkomargo@gmail.com}}
\vskip 3mm
\address{
${}^a$Physics and Astronomy Department,\\ Stony Brook University, Stony Brook, NY 11794, USA
}
\address{
${}^b$Perimeter Institute for Theoretical Physics,\\
Waterloo, Ontario, N2L 2Y5, Canada}
\address{
${}^c$ Department of Physics, University of Waterloo,\\ Waterloo, ON N2L 3G1, Canada}
\address{
${}^d$ Simons Center for Geometry and Physics,\\ Stony Brook University, Stony Brook, NY 11794, USA}
\address{
${}^e$ Weizmann Institute of Science,\\ Rehovot, Israel}

\end{center}

\vspace{5mm}
\abstract{
\noindent

Three-dimensional gauge theories coupled to fermions can develop interesting nonperturbative dynamics. Here we study in detail the dynamics of $SU(N)$ gauge theories coupled to a Dirac fermion in the rank-two symmetric and antisymmetric representation. We argue that when the Chern-Simons level is sufficiently small the theory develops a quantum phase with an emergent topological field theory. When the Chern-Simons level vanishes, we further argue that a baryon condenses and hence baryon symmetry is spontaneously broken. The infrared theory then consists of a Nambu-Goldstone boson coupled to a topological field theory. Our proposals also lead to new fermion-fermion dualities involving fermions in two-index representations. We make contact between our proposals and some recently discussed aspects of four-dimensional gauge theories. This leads us to a proposal for the domain wall theories of non-supersymmetric gauge theories with fermions in two-index representations. Finally, we discuss some  aspects of the time-reversal anomaly in theories with a one-form symmetry.}
\vfill

\end{titlepage}

{\hypersetup{linkcolor=black}
\tableofcontents
\thispagestyle{empty}
}

\section{Introduction}
\label{sec:intro}
\setcounter{footnote}{0}

Strongly coupled Quantum Field Theories (QFTs) can develop interesting infrared ``quantum" (nonperturbative) phases, distinct from those that can be inferred by semiclassical considerations. Recently the study of the infrared dynamics of $2+1$ dimensional gauge theories has resulted in the discovery of novel nonperturbative quantum phases~\cite{Komargodski:2017keh,Gomis:2017ixy,Cordova:2017vab,Bashmakov:2018wts,Benini:2018umh, Choi:2018ohn}. In addition, some related aspects have been already studied on the lattice~\cite{Karthik:2016bmf,Karthik:2018nzf}. The subject involves several conceptual differences from the (perhaps) more familiar setting of $3+1$ dimensional gauge theories coupled to matter: 

\begin{itemize}

\item Since the gauge coupling has positive mass dimension, $2+1$ dimensional gauge theories are always asymptotically free. In particular, these theories are interesting even when the gauge group is $U(1)$. 

\item $2+1$ dimensional gauge theories are labeled by a gauge group, matter fields, and the Chern-Simons couplings, which are quantized. 

\item Since in $2+1$ dimensions there is no notion of spinor chirality, one can typically add a mass term for the matter fields  preserving all the continuous symmetries of the massless theory. In some theories, however, the massless point is distinguished by the presence of (a discrete) antilinear time-reversal symmetry. The phases of these theories can be studied as a function of the continuous mass deformations for the matter fields.

\item There are no continuous 't Hooft anomalies for the symmetries of $2+1$ dimensional theories. However, there are many discrete anomalies and they have to be consistent with the infrared phases of these theories. Even though these anomalies are discrete they nevertheless provide highly nontrivial constraints on the infrared dynamics. In addition, such quantum phases are constrained by the matching of some counterterms. This matching of counterterms physically corresponds to consistency conditions on conductivity coefficients. We will encounter some examples in this paper. 

\end{itemize}

The subject of $2+1$ dimensional gauge theories connects in an obvious way to condensed matter physics (where the gauge symmetry is typically emergent) and in somewhat less obvious ways to particle physics. Many $3+1$ dimensional gauge theories have degenerate trivial vacua. As a result, a domain wall connecting two vacua supports at low energies a $2+1$ dimensional theory, and one can often make this connection very natural (as we will see in this paper). In addition, one can study $3+1$ dimensional gauge theories compactified on a circle (e.g.~in the context of finite temperature physics), a problem that similarly reduces to the study of $2+1$ dimensional systems.

In spite of the fact that $2+1$ dimensional gauge theories are asymptotically free, these theories admit regimes in parameter space where they are weakly coupled and the infrared dynamics can be inferred by a careful semiclassical analysis:

\begin{itemize}

\item When all the matter fields have a large mass (in units of the Yang-Mills coupling constant) they can be integrated out at energy scales above the strong coupling scale. This only leads to a shift in the infrared Chern-Simons level (this shift is one-loop exact for fermions and trivial for bosons) and hence the deep infrared theory is given by the infrared dynamics of the pure Yang-Mills-Chern-Simons gauge theory, which is quite well understood. This typically leads in the infrared to a Chern-Simons Topological Quantum Field Theory (TQFT).\footnote{An exception to this is when the Chern-Simons level vanishes upon integrating out massive matter fields. In that case, for simply connected groups, the TQFT is trivial. For non-simply connected groups the situation is more complicated, but we will not need it here except in the case of $U(1)$, where the low energy theory is the gapless theory of a compact scalar. This fact will play an important role in this paper.} Since the shift of the Chern-Simons level depends on the sign of the mass of the fermion~\cite{Niemi:1983rq,Redlich:1983kn,Redlich:1983dv}, the large mass limit  defines an asymptotically large-positive mass phase and an asymptotically large-negative mass phase, described  by two distinct TQFTs.

\item When the number of matter fields is very large\footnote{The relevant group theory factor is the index of the (possibly reducible) representation of the matter fields.}~\cite{Appelquist:1988sr,Appelquist:1989tc} (i.e.~many species or large representations) one can demonstrate that there exists a weakly coupled conformal field theory (CFT) which interpolates between the  TQFTs describing the two asymptotically large mass phases. Such a theory does not develop interesting new quantum phases.

\item Likewise, when the Chern-Simons level is very large~\cite{Avdeev:1991za} one can show that there exists a weakly coupled  CFT  interpolating between the  TQFTs describing the two asymptotically large mass phases. For large $k$ the theory does not develop interesting new quantum phases. 

\smallskip

While no new quantum phases emerge for ``large representations" or  large Chern-Simons level, it is a wide-open nonperturbative problem to determine for which representations and which levels new quantum phases develop. For some recent work on such questions in the context of quiver gauge theories see~\cite{Jensen:2017dso,Aitken:2018cvh} and references therein.
 
\end{itemize}

It follows from our discussion above that the dynamics of $2+1$ dimensional gauge theories is especially interesting when neither the Chern-Simons level, the dimension of the representation, nor the mass of the matter fields are too large. In this regime there is no semiclassical approximation to the dynamics of the theory. This is when we may expect quantum effects to dominate the dynamics and new interesting phenomena may emerge, including new nonperturbative phases. 

In this paper we study a class of $2+1$ dimensional gauge theories for which we provide a large body of evidence that they indeed develop new nonperturbative phases along with new phase transitions, for which we propose novel dual descriptions. (We do not know, in general, if these phase transitions are 1st or 2nd order.) The theories we analyze are $SU(N)$ Yang-Mills gauge theories coupled to a  fermion in the rank-two symmetric or antisymmetric representation of $SU(N)$ and a Chern-Simons term at level $k$. (The fermion is a Dirac fermion, i.e.~a complex fermion with two components.)

For generic $k$, these models have a global baryon number symmetry, $U(1)_B$, acting on the fermion as $\psi\mapsto e^{i\alpha}\psi$. In addition, there is charge-conjugation symmetry. Both of these symmetries are unbroken by the mass term $im\bar\psi\psi$.  For $k=0$ (which is only allowed for even $N$) the model also admits a time-reversal symmetry. The mass term breaks the time-reversal symmetry. Finally, since there are no dynamical degrees of freedom in the fundamental representation, these theories have a one-form $\mathbb{Z}_2$ symmetry when $N$ is even. (In the context of condensed matter physics, the one-form symmetry is expected to be accidental.)

\vspace{8pt}

Let us now summarize the main results:

\begin{enumerate}

\item These theories have a critical value of the level $k_\mathrm{crit}$ below which a new intermediate quantum phase appears between the semiclassically accessible asymptotic large-positive and large-negative mass phases. The critical value is\footnote{The quantized level $k$ must be integer for $N$ even and half-integer for $N$ odd. See section~\ref{sec:phases}.}
\begin{equation}
\text{symmetric:}\ k_\mathrm{crit}=\frac{N+2}{2}\,, \qquad \qquad \text{antisymmetric:}\ k_\mathrm{crit}=\frac{N-2}{2}\,.
\end{equation}

\item For $0\neq k<k_\mathrm{crit}$ there is an intermediate quantum phase described by the following ``emergent'' TQFTs\footnote{We recall that $U(N)_{P,Q}:= \frac{SU(N)_P\times U(1)_{NQ}}{\mathbb Z_N}$ with $P\equiv Q~\text{mod}~N$. The quotient by $\mathbb Z_N$ gauges an anomaly-free one-form symmetry~\cite{Kapustin:2014gua,Gaiotto:2014kfa}.}
\begin{equation}
\text{symmetric:}\ U\!\left(\frac{N+2}{2}-k\right)_{\frac{N+2}{2}+k,2k} \,, ~ ~ \text{antisymmetric:}\ U\!\left(\frac{N-2}{2}-k\right)_{\frac{N-2}{2}+k,2k}\,.
\label{intermeda}
\end{equation}

\item For $k=0$ there is  an intermediate quantum phase that includes a Nambu-Goldstone boson (NGB) corresponding to the spontaneous breaking of $U(1)_B$, along with a TQFT:
\begin{equation}
\text{symmetric:}\ \frac{SU\!\left(\frac{N+2}{2}\right)_{\frac{N+2}{2}}\times S^1}{\mathbb Z_{\frac{N+2}{2}}} \,, ~ ~ \text{antisymmetric:}\ \frac{SU\!\left(\frac{N-2}{2}\right)_{\frac{N-2}{2}}\times S^1}{\mathbb Z_{\frac{N-2}{2}}}\,.
\end{equation}
The notation $S^1$ stands for the linear sigma model of a compact real scalar field $\{\phi\sim \phi+2\pi\}$, which is dual to pure $U(1)_0$ gauge theory in $2+1$ dimensions.\footnote{For $k=0$, the $U(1)_0$ factor in~\eqref{intermeda} should not be interpreted as a TQFT, but rather as a gapless $U(1)_0$ gauge theory, which can be dualized to the compact scalar.} The scalar couples to the TQFT $SU\!\left(\frac{N\pm 2}{2}\right)_{\frac{N\pm 2}{2}}$ by gauging a diagonal, anomaly-free $\mathbb Z_{\frac{N\pm 2}{2}}$ one-form symmetry.

The microscopic (ultraviolet) theory for $k=m=0$ is time-reversal invariant. It may then seem odd that we are proposing that this model flows to a TQFT coupled to a Nambu-Goldstone boson, as TQFTs are typically non-time-reversal invariant. It is encouraging to observe that the $SU(n)_n/\mathbb Z_n$ Chern-Simons theory is in fact also a time-reversal invariant (spin) TQFT~\cite{Aharony:2016jvv}. This is a  nontrivial consistency check of our proposal.

\item For $k<k_\mathrm{crit}$ these theories undergo  two phase transitions as a function of the mass of the fermion. The phase transitions connect the intermediate quantum phase with the asymptotic large-positive mass phase and with 
the asymptotic large-negative mass phase, respectively. We propose that these transitions have a dual description in terms of another $2+1$ dimensional gauge theory. This leads us to propose new (fermion-fermion) dualities in $2+1$ dimensions.

 \smallskip
Dualities for $SU(N)_k+~\text{symmetric}~ \psi$ for $k<\frac{N+2}{2}$:
\begin{equation}
\begin{aligned}
SU(N)_k+~\text{symmetric}~ \psi &\longleftrightarrow U\!\left(\frac{N+2}{2}+k\right)_{\frac12(-1+k-3N/2),k-N/2}+~\text{antisymmetric}~ \tilde\psi\\[+5pt]
SU(N)_k+~\text{symmetric}~ \psi &\longleftrightarrow U\!\left(\frac{N+2}{2}-k\right)_{\frac12(+1+k+3N/2),k+N/2}+~\text{antisymmetric}~ \hat\psi\,.
\end{aligned}
\end{equation}

\medskip

Dualities for $SU(N)_k+~\text{antisymmetric}~ \psi$ for $k<\frac{N-2}{2}$:
\begin{equation}
\begin{aligned}
SU(N)_k+~\text{antisymmetric}~ \psi &\longleftrightarrow U\!\left(\frac{N-2}{2}+k\right)_{\frac12(+1+k-3N/2),k-N/2}+~\text{symmetric}~ \tilde\psi\\[+5pt]
SU(N)_k+~\text{antisymmetric}~ \psi &\longleftrightarrow U\!\left(\frac{N-2}{2}-k\right)_{\frac12(-1+k+3N/2),k+N/2}+~\text{symmetric}~ \hat\psi\,.
\end{aligned}
\end{equation}
We note that the fermion in the dual gauge theory transforms in the other rank-two representation compared to the  fermion in the original gauge theory.

\item For $k\geq k_\mathrm{crit}$ the phase diagram has just two phases: the asymptotic large-positive mass and large-negative mass semiclassical phases, separated by a phase transition. For very large $k$ the phase transition is controlled by a weakly coupled CFT. The asymptotic large mass phases are the TQFTs
\begin{equation}
\text{symmetric:}\ SU\!\left(N\right)_{k\pm \frac{N+2}{2} } \,, \qquad\text{antisymmetric:}\ SU\!\left(N\right)_{k\pm \frac{N-2}{2}}\,,
\label{intermed}
\end{equation}
where the upper/lower sign is for the large positive/negative mass asymptotic phase. These phases are present in these theories for any $k$, but for $k<k_\mathrm{crit}$ they are separated by the intermediate quantum phases discussed above while for $k\geq k_\mathrm{crit}$ they are separated by a single transition (which for sufficiently large $k$ must be given by a CFT). 

\end{enumerate}

{\begin{center}\underline{Outline of the Paper} \end{center}}
In section 2 we present our conjectures for the phases of $SU(N)$ Yang-Mills gauge theory coupled to a fermion in the symmetric and antisymmetric representation  and with a Chern-Simons term. In section 3 we present a list of nontrivial consistency checks, involving a comparison to known special cases, and the highly nontrivial matching of some contact terms. In section 4 we study some domain wall solutions in four-dimensional gauge theories and show that at least in one case one can explicitly demonstrate baryon symmetry breaking on the domain wall, in agreement with the predictions for the infrared behavior of the corresponding three-dimensional gauge theory.  In section 5 we make a few forward-looking comments about  the time-reversal anomaly and about baryons.

\section{Phase Diagrams}
\label{sec:phases}

Consider Yang-Mills theory with gauge group $SU(N)$, a Chern-Simons term and a Dirac fermion $\psi$ in the rank-two symmetric or antisymmetric representation $R$ of $SU(N)$:
\begin{equation}\label{theoryL}
\mathcal L= \Tr\left(-\frac{1}{2g^2}F^2 +\frac{ik_\mathrm{bare}}{4\pi} \left(A\mathrm dA+\frac23 A^3 \right) +i\bar \psi \slashed{D} \psi+i m \bar\psi \psi\right)\,.
\end{equation}

In this section we make a proposal for the phase diagram of these theories as a function of the effective Chern-Simons level $k:=k_\mathrm{bare}-T(R)$ and of the  mass $m\in \mathbb R$ of the fermion.\footnote{The shift of the Chern-Simons level in~\eqref{theoryL} by $T(R)$ arises from the determinant of the massless fermion. It is convenient to use $k_\mathrm{bare}$ when writing Lagrangians since $k_\mathrm{bare}$ is always an integer. However, the infrared phases of the theory are more conveniently labeled by $k:=k_\mathrm{bare}-T(R)$ because time-reversal symmetry acts on $k$ by simply reversing it, along with reversing the mass of the fermion.} $T(R)$ is the Dynkin index of the $SU(N)$ representation under which the Dirac fermion transforms. Since under time-reversal $k\to -k$ (along with reversing the sign of the mass), we restrict our discussion to $k\geq 0$. Since $k_\mathrm{bare}\in \mathbb Z$, it follows from table~\ref{table1} that $k\in \mathbb Z$ for $N$ even and $k\in \mathbb Z+\frac12$ for $N$ odd.

\begin{table}[!h]
\begin{equation*}\arraycolsep=6pt\def\arraystretch{1.4}
\begin{array}{|c|c|c|c|}\hline
& \text{symmetric} & \text{antisymmetric} & \text{adjoint} \\ \hline
T(R) & \frac12(N+2) & \frac12(N-2) & N\\\hline
\end{array}
\end{equation*}
\caption{Index for $SU(N)$ rank-two  and adjoint representations. (Since the adjoint representation is real, one could also take the corresponding fermion to be a Majorana fermion -- that model was discussed in detail in~\cite{Gomis:2017ixy}.)} \label{table1}
\end{table}

We now discuss the global symmetries of these theories. There is a $U(1)_B$ flavor symmetry and a $\mathbb Z^\mathsf C_2$ charge-conjugation symmetry $\mathsf C$ acting as\footnote{For $N=2$ the action of $\mathsf C$ on the gauge field is a gauge transformation.} 
\begin{equation}\label{actsymm}
U(1)_B:~ \psi\mapsto e^{i\alpha} \psi\,,\qquad \qquad \mathsf C:~\begin{cases}A_\mu \mapsto -A_\mu^T  &  \\[+2pt]
\, \psi\  \mapsto +\psi^* \,.\end{cases}
\end{equation}

These transformations do not commute and generate the group $O(2)_B=U(1)_B\rtimes \mathbb Z^\mathsf C_2$. Since the center of the gauge group $\mathbb Z_N\subset SU(N)$ acts as $\psi\mapsto e^{\frac{4\pi i}{N}} \psi$, the  global symmetry group is $O(2)_B/\mathbb Z_{N/2}$ for $N$ even and $O(2)_B/\mathbb Z_{N}$ for $N$ odd. The operators charged under this symmetry are baryons, which will be discussed briefly at the end of this paper.

Since the gauge group is simply-connected, the magnetic symmetry group is trivial. For $N$ even a $\mathbb Z_2\subset \mathbb Z_N$ subgroup of the center acts trivially on $\psi$ and the theory has a $\mathbb Z_2$ one-form global symmetry. For $N$ odd the one-form symmetry is trivial. Finally, for $k=m=0$ the theory is time-reversal invariant. 
Time-reversal symmetry acts on the fermion by 
\begin{equation}
\mathsf T :~ \psi \mapsto \gamma_0\psi^*~. 
\end{equation}
Therefore, in these theories, $\mathsf T^2=(-1)^F$. It is easy to verify that the mass term is odd under time-reversal symmetry  but it preserves all other symmetries.

We proceed now to analyzing the phase diagram. We start with the phases that can be established by a semiclassical analysis. When $|m|\gg g^2$ we can reliably integrate out the fermion before the interactions become strong. Integrating out a massive fermion shifts the Chern-Simons level to $k+\sign(m) T(R)$, and the resulting effective theory is pure $SU(N)$ Yang-Mills with an integer-quantized Chern-Simons term at level $k+\sign(m) T(R)$. This theory, which now has no matter fields, flows at low energies to the topological $SU(N)$ Chern-Simons theory at level $k+\sign(m) T(R)$, which we denote by $SU(N)_{k+\sign(m) T(R)}$. Therefore the infrared dynamics is captured by the TQFTs $SU(N)_{k\pm T(R)}$ for large positive and negative mass respectively. These asymptotic large mass phases are present for all $k$ and all $N$. In the above discussion of the asymptotic phases, $k=T(R)$ is an exception since the infrared theory (after integrating the fermions out) is pure Yang-Mills theory (without a Chern-Simons term) with gauge group $SU(N)$. In this case the infrared theory is trivial and gapped due to confinement and due to the fact that the gauge group is simply-connected.

\begin{figure}[!h]
\centering
\begin{tikzpicture}
\node[scale=1.1] at (5.5,3) {$SU(N)_k+\ydiagram{2}\,\psi$};
\node at (13,3) {$k\ge\frac{N+2}{2}$};
\draw (4.1,2.65) -- (6.9,2.65);
\draw[thick,<->,>=stealth] (-1.5,0) -- (13.5,0);
\filldraw[white!20!blue] (5.7,0) circle (3pt);\draw (5.7,0) circle (3pt);

\node[scale=1.1] at (2,-.8) {$SU(N)_{k-\frac12(N+2)}$};
\node[scale=1.1] at (9.5,-.8) {$SU(N)_{k+\frac12(N+2)}$};

\node at (13,.4) {\footnotesize$m\to+\infty$};
\node at (-1,.4) {\footnotesize$m\to-\infty$};
\end{tikzpicture}
\caption{Phase diagram of $SU(N)$ gauge theory with a symmetric fermion for $k\geq\frac{N+2}{2}$. The solid circle represents a phase transition between the asymptotic phases. For sufficiently large $k$ we know for certain that the phase transition is associated with a CFT.}
\label{symmlargek}
\end{figure}
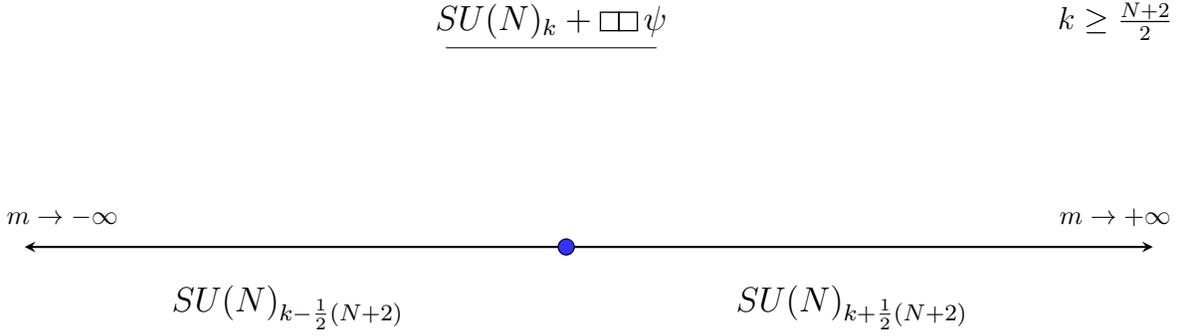

\begin{figure}[!h]
\centering
\begin{tikzpicture}
\node[scale=1.1] at (5.5,3) {$SU(N)_k+\ydiagram{1,1}\,\psi$};
\node at (13,3) {$k\ge\frac{N-2}{2}$};
\draw (4.1,2.65) -- (6.9,2.65);
\draw[thick,<->,>=stealth] (-1.5,0) -- (13.5,0);
\filldraw[white!20!blue] (5.7,0) circle (3pt);\draw (5.7,0) circle (3pt);

\node[scale=1.1] at (2,-.8) {$SU(N)_{k-\frac12(N-2)}$};
\node[scale=1.1] at (9.5,-.8) {$SU(N)_{k+\frac12(N-2)}$};

\node at (13,.4) {\footnotesize$m\to+\infty$};
\node at (-1,.4) {\footnotesize$m\to-\infty$};
\end{tikzpicture}
\caption{Phase diagram of $SU(N)$ with an antisymmetric fermion for $k\geq\frac{N-2}{2}$. The solid circle represents a phase transition between the asymptotic phases. For sufficiently large $k$ we know for certain that the phase transition is associated with a CFT.}
\label{asymmlargek}
\end{figure}
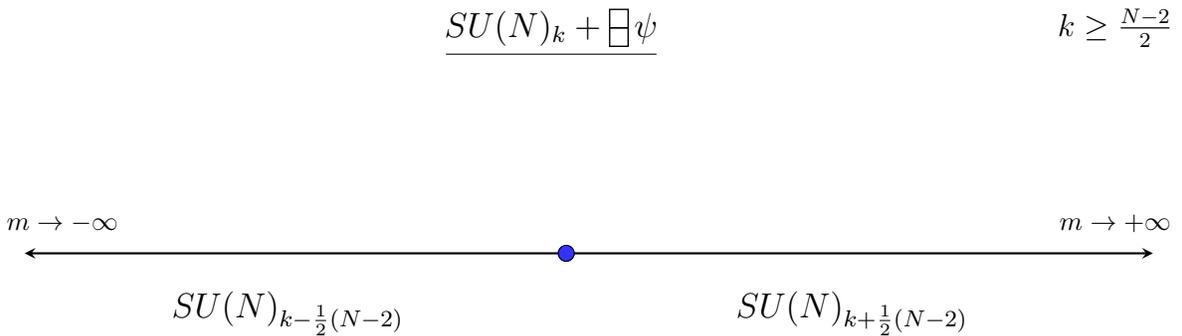

As long as $k$ is sufficiently large, the above two topological phases $SU(N)_{k+\sign(m) T(R)}$ are separated by a single transition. The question is below which value of $k$ additional phases appear. Our proposal is that as long as $k\geq T(R)$ the above picture holds true, namely, the two semiclassically accessible phases are separated by a phase transition at some value of the mass. The phase diagrams for $k\geq T(R)$ are summarized in figures~\ref{symmlargek} and~\ref{asymmlargek}. Note that the boundary of the above region, $k=T(R)$, is exactly where one of the phases becomes a trivial infrared theory, with no Chern-Simons TQFT.

\begin{figure}[!h]
\centering
\begin{tikzpicture}
\node[scale=1.1] at (5.5,4) {$SU(N)_k+\ydiagram{1,1}\,\psi$};
\node at (13,4) {$0<k<\frac{N-2}{2}$};
\draw (4.1,3.65) -- (6.9,3.65);
\draw[thick,<->,>=stealth] (-1.5,0) -- (13.5,0);
\filldraw[white!20!blue] (3.2,0) circle (3pt);\draw (3.2,0) circle (3pt);
\filldraw[white!20!blue] (8+.35,0) circle (3pt);\draw (8+.35,0) circle (3pt);
\node[scale=1.1] at (1.7,2) {$U\!\left(\frac{N-2}{2}-k\right)_{\frac14(3N+2k-2),k+\frac12N}+\ydiagram{2}\,\hat\psi$};
\node[scale=1.1] at (10.1,2) {$ U\!\left(\frac{N-2}{2}+k\right)_{\frac14(-3N+2k+2),k-\frac12N}+\ydiagram{2}\,\tilde\psi$};
\draw[thick,->,>=stealth] (2.4,1.5) -- (3.1,.2);
\draw[thick,->,>=stealth] (8.8+.35,1.5) -- (8.1+.35,.2);

\node[scale=1.1] at (.85,-.8) {$SU(N)_{k-\frac12(N-2)}$};
\draw[thick,<->,>=stealth] (.85,-1.3) -- (.85,-2.2);
\node[scale=1.1] at (.85,-2.7) {$U\!\left(\frac{N-2}{2}-k\right)_{+N}$};
\node[scale=1.1] at (5.7+.075,-.8) {$ U\!\left(\frac{N-2}{2}-k\right)_{+\frac12(N-2)+k,2k}$};
\draw[thick,<->,>=stealth] (5.7+.075,-1.3) -- (5.7+.075,-2.2);
\node[scale=1.1] at (5.7+.075,-2.7) {$U\!\left(\frac{N-2}{2}+k\right)_{-\frac12(N-2)+k,2k}$};
\node[scale=1.1] at (10.925,-.8) {$SU(N)_{k+\frac12(N-2)}$};
\draw[thick,<->,>=stealth] (10.925,-1.3) -- (10.925,-2.2);
\node[scale=1.1] at (10.925,-2.7) {$U\!\left(\frac{N-2}{2}+k\right)_{-N}$};

\node at (13,.4) {\footnotesize$m\to+\infty$};
\node at (-1,.4) {\footnotesize$m\to-\infty$};
\end{tikzpicture}
\caption{Phase diagram of $SU(N)$ with an antisymmetric fermion for $0<k<\frac{N-2}{2}$.  The solid circles represent a phase transition between the asymptotic phases and the intermediate quantum phase. Each phase transition has a dual gauge theory description, which appears with an arrow pointing to the phase transition. The mass deformations are related by 
$m_\psi=-m_{\hat\psi}$ and $m_\psi=-m_{\tilde\psi}$.}
\label{asymmsmallk}
\end{figure}
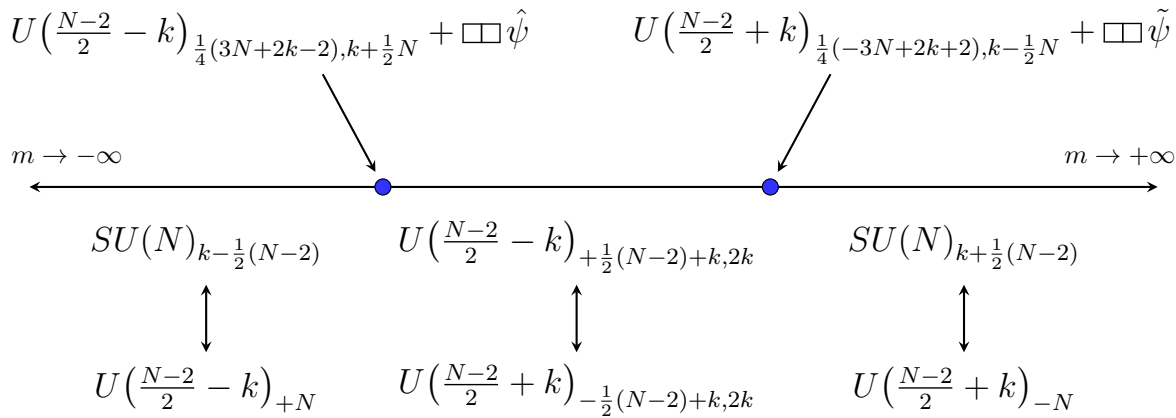

For $0\leq k<T(R)$ we propose that there is a new intermediate ``quantum phase" in between the asymptotic large mass phases.\footnote{In the case of symmetric fermion the new phase appears at $k=T(R)-2= \frac{N}{2}-1$. See below.} This phase is inherently quantum mechanical, and  is not visible semiclassically. This new quantum phase  connects to each of the asymptotic phases through a phase transition. The phase diagrams for $0< k<T(R)$ are summarized in figures~\ref{symmsmallk} and~\ref{asymmsmallk}.  The reason that $k=0$ is excluded from the figures is that it requires a separate discussion, as we shall see below.

\begin{figure}[!h]
\centering
\begin{tikzpicture}
\node[scale=1.1] at (5.5,4) {$SU(N)_k+\ydiagram{2}\,\psi$};
\node at (13,4) {$0< k<\frac{N+2}{2}$};
\draw (4.1,3.65) -- (6.9,3.65);
\draw[thick,<->,>=stealth] (-1.5,0) -- (13.5,0);
\filldraw[white!20!blue] (3.2,0) circle (3pt);\draw (3.2,0) circle (3pt);
\filldraw[white!20!blue] (8+.35,0) circle (3pt);\draw (8+.35,0) circle (3pt);
\node[scale=1.1] at (1.7,2) {$ U\!\left(\frac{N+2}{2}-k\right)_{\frac14(3N+2k+2),k+\frac12N}+\ydiagram{1,1}\,\hat\psi$};
\node[scale=1.1] at (10.1,2) {$U\!\left(\frac{N+2}{2}+k\right)_{\frac14(-3N+2k-2),k-\frac12N}+\ydiagram{1,1}\,\tilde\psi$};
\draw[thick,->,>=stealth] (2.4,1.5) -- (3.1,.2);
\draw[thick,->,>=stealth] (8.8+.35,1.5) -- (8.1+.35,.2);

\node[scale=1.1] at (.85,-.8) {$SU(N)_{k-\frac12(N+2)}$};
\draw[thick,<->,>=stealth] (.85,-1.3) -- (.85,-2.2);
\node[scale=1.1] at (.85,-2.7) {$U\!\left(\frac{N+2}{2}-k\right)_{+N}$};
\node[scale=1.1] at (5.7+.075,-.8) {$U\!\left(\frac{N+2}{2}-k\right)_{+\frac12(N+2)+k,2k}$};
\draw[thick,<->,>=stealth] (5.7+.075,-1.3) -- (5.7+.075,-2.2);
\node[scale=1.1] at (5.7+.075,-2.7) {$U\!\left(\frac{N+2}{2}+k\right)_{-\frac12(N+2)+k,2k}$};
\node[scale=1.1] at (10.925,-.8) {$SU(N)_{k+\frac12(N+2)}$};
\draw[thick,<->,>=stealth] (10.925,-1.3) -- (10.925,-2.2);
\node[scale=1.1] at (10.925,-2.7) {$U\!\left(\frac{N+2}{2}+k\right)_{-N}$};

\node at (13,.4) {\footnotesize$m\to+\infty$};
\node at (-1,.4) {\footnotesize$m\to-\infty$};
\end{tikzpicture}
\caption{Phase diagram of $SU(N)$ with a symmetric fermion for $0<k<\frac{N+2}{2}$.  The solid circles represent a phase transition between the asymptotic phases and the intermediate quantum phase. Each phase transition has a dual gauge theory description, which appears with an arrow pointing to the phase transition. The mass deformations are related by 
$m_\psi=-m_{\hat\psi}$ and $m_\psi=-m_{\tilde\psi}$.}
\label{symmsmallk}
\end{figure}
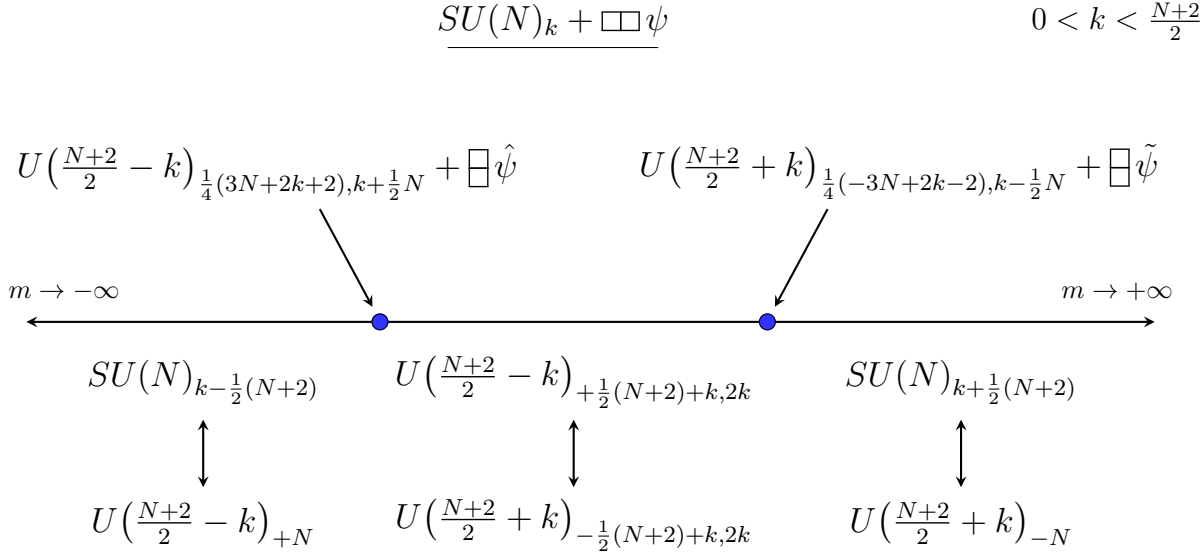

The way we arrive at the phase diagrams in figures~\ref{symmsmallk} and~\ref{asymmsmallk} is as follows. As mentioned above the asymptotic positive and negative mass phases are described by the TQFTs $SU(N)_{k\pm T(R)}$. These TQFTs admit a level/rank $SU/U$ dual description~\cite{Nakanishi:1990hj,Hsin:2016blu}\footnote{Level/rank dualities are generically valid only as spin TQFTs, and therefore, whenever the theory on one side of the duality is not spin (i.e.~it does not have a transparent spin $1/2$ line) we must tensor that theory with a trivial spin TQFT.
$SU(N)_k$ is never spin and $U(N)_{k,k}$ is spin for $k$ odd.}
\begin{equation}
SU(N)_{k\pm T(R)}\ \longleftrightarrow\ U(T(R)\pm k)_{\mp N,\mp N}\,.
\end{equation}

We start with the level/rank dual description of the asymptotic positive mass phase and search for a dual description that would allow us to understand the quantum phase semiclassically in the dual variables. Similarly, we consider the asymptotic negative mass phase and search for an ultraviolet gauge theory that could describe the phase transition between the semiclassical asymptotic negative mass phase and the quantum phase. These steps involve some guesswork. The fact that this can be done at all is already a highly nontrivial consistency check. Indeed, the two required dual descriptions are mutually non-local\footnote{By ``mutually non-local'' we mean that there exists no local map between the fields in the two dual descriptions, yet, they have some region of overlap in the physics they describe. This is reminiscent of the Seiberg-Witten solution, which has two mutually non-local theories describing two different massless theories, with a region of overlap.} but there is only one quantum phase, which both of them have to describe simultaneously (in our case this happens thanks to the new level/rank duality in~\cite{Hsin:2016blu}). In the present context, luckily, we were able to find consistent dual descriptions describing the same quantum phase. Furthermore, this guess satisfies very nontrivial additional consistency checks, as we shall see. 
One of the dual theories is based on the gauge group $U(T(R)- k)$ and the other on the gauge group $U(T(R)+ k)$ with appropriate Chern-Simons levels and matter representations.

We are thus led to propose the following new fermion-fermion dualities for $0\leq k<T(R)$:
\medskip

\noindent
Dualities for $SU(N)_k+~\text{symmetric}~ \psi$ for $k<\frac{N+2}{2}$:
\begin{equation}
\begin{aligned}
SU(N)_k+~\text{symmetric}~ \psi &\longleftrightarrow U\!\left(\frac{N+2}{2}+k\right)_{\frac12(-1+k-3N/2),k-N/2}+~\text{antisymmetric}~ \tilde\psi\\[+5pt]
SU(N)_k+~\text{symmetric}~ \psi &\longleftrightarrow U\!\left(\frac{N+2}{2}-k\right)_{\frac12(+1+k+3N/2),k+N/2}+~\text{antisymmetric}~ \hat\psi\,.
\label{symmdual}
\end{aligned}
\end{equation}

\medskip

\noindent
Dualities for $SU(N)_k+~\text{antisymmetric}~ \psi$ for $k<\frac{N-2}{2}$:
\begin{equation}
\begin{aligned}
SU(N)_k+~\text{antisymmetric}~ \psi &\longleftrightarrow U\!\left(\frac{N-2}{2}+k\right)_{\frac12(+1+k-3N/2),k-N/2}+~\text{symmetric}~ \tilde\psi\\[+5pt]
SU(N)_k+~\text{antisymmetric}~ \psi &\longleftrightarrow U\!\left(\frac{N-2}{2}-k\right)_{\frac12(-1+k+3N/2),k+N/2}+~\text{symmetric}~ \hat\psi\,.
\end{aligned}
\end{equation}

For $k=T(R)-1=N/2$ in the theory of the fermion in the symmetric representation the intermediate phase coincides with the asymptotic large negative mass phase and the first phase transition is therefore unnecessary. Indeed, the associated duality in the second line of~\eqref{symmdual} trivializes since the antisymmetric representation of $U(1)$ is trivial.
 
The case of $k=0$ is particularly interesting and requires a separate discussion. The quantum phase that has appeared in the figure~\ref{symmsmallk} and~\ref{asymmsmallk} would seem to make sense also for $k=0$. However, while for $k>0$ it is a pure TQFT, for $k=0$ it is not. Indeed, after integrating the fermion in the dual theory with gauge group $U(T(R))$, we are left with pure $U(T(R))_{T(R),0}$ Yang-Mills-Chern-Simons theory, with $T(R)=\frac N2 \pm 1$ in the symmetric/antisymmetric case. The crucial point is that the $U(1)_0$ factor is not topological. This latter theory can be dualized to the theory of a compact, real scalar field $\phi$
\begin{equation}\label{compact}
\mathcal L_{U(1)_0}=\frac{f_\pi^2}{2}(\partial\phi)^2\,.
\end{equation}
with $\phi\sim \phi+2\pi$ and $f_\pi^2$ the ``decay constant''. 

\begin{figure}[!h]
\centering
\begin{tikzpicture}
\node[scale=1.1] at (5.5,4) {$SU(N)_0+\ydiagram{2}\,\psi$};
\draw (4.1,3.65) -- (6.9,3.65);
\draw[thick,<->,>=stealth] (-1.5,0) -- (13.5,0);
\filldraw[white!20!blue] (3.2,0) circle (3pt);\draw (3.2,0) circle (3pt);
\filldraw[white!20!blue] (8+.35,0) circle (3pt);\draw (8+.35,0) circle (3pt);
\node[scale=1.1] at (1.7,2) {$U\!\left(\frac{N+2}{2}\right)_{\frac14(3N+2),\frac12N}+\ydiagram{1,1}\,\psi$};
\node[scale=1.1] at (10.1,2) {$ U\!\left(\frac{N+2}{2}\right)_{-\frac14(3N+2),-\frac12N}+\ydiagram{1,1}\,\psi$};
\draw[thick,->,>=stealth] (2.4,1.5) -- (3.1,.2);
\draw[thick,->,>=stealth] (8.8+.35,1.5) -- (8.1+.35,.2);

\node[scale=1.1] at (.85,-.8) {$SU(N)_{-\frac12(N+2)}$};
\draw[thick,<->,>=stealth] (.85,-1.3) -- (.85,-2.2);
\node[scale=1.1] at (.85,-2.7) {$U\!\left(\frac{N+2}{2}\right)_{+N}$};
\node at (5.7+.075,-.8) {$\dfrac{SU\!\left(\frac{N+2}{2}\right)_{+\frac12(N+2)}\times S^1}{\mathbb Z_{\frac12(N+2)}}$};
\draw[thick,<->,>=stealth] (5.7+.075,-1.5) -- (5.7+.075,-2);
\node at (5.7+.075,-2.7) {$\dfrac{SU\!\left(\frac{N+2}{2}\right)_{-\frac12(N+2)}\times S^1}{\mathbb Z_{\frac12(N+2)}}$};
\node[scale=1.1] at (10.925,-.8) {$SU(N)_{\frac12(N+2)}$};
\draw[thick,<->,>=stealth] (10.925,-1.3) -- (10.925,-2.2);
\node[scale=1.1] at (10.925,-2.7) {$U\!\left(\frac{N+2}{2}\right)_{-N}$};

\node at (5.7+.075,1) {\footnotesize$m=0$};
\draw[thick,->,>=stealth] (5.7+.075,.7) -- (5.7+.075,.2);
\filldraw (5.7+.075,0) circle (1pt);
\node at (13,.4) {\footnotesize$m\to+\infty$};
\node at (-1,.4) {\footnotesize$m\to-\infty$};
\end{tikzpicture}
\caption{Phase diagram of $SU(N)$ gauge theory with a symmetric fermion for $k=0$. The circle $S^1$ represents the corresponding sigma model. Each phase transition has a dual gauge theory description, which appears with an arrow pointing to the phase transition. The mass deformations are related by 
$m_\psi=-m_{\hat\psi}$ and $m_\psi=-m_{\tilde\psi}$.}
\label{fig:phase_diag_symm_zero_k}
\end{figure}
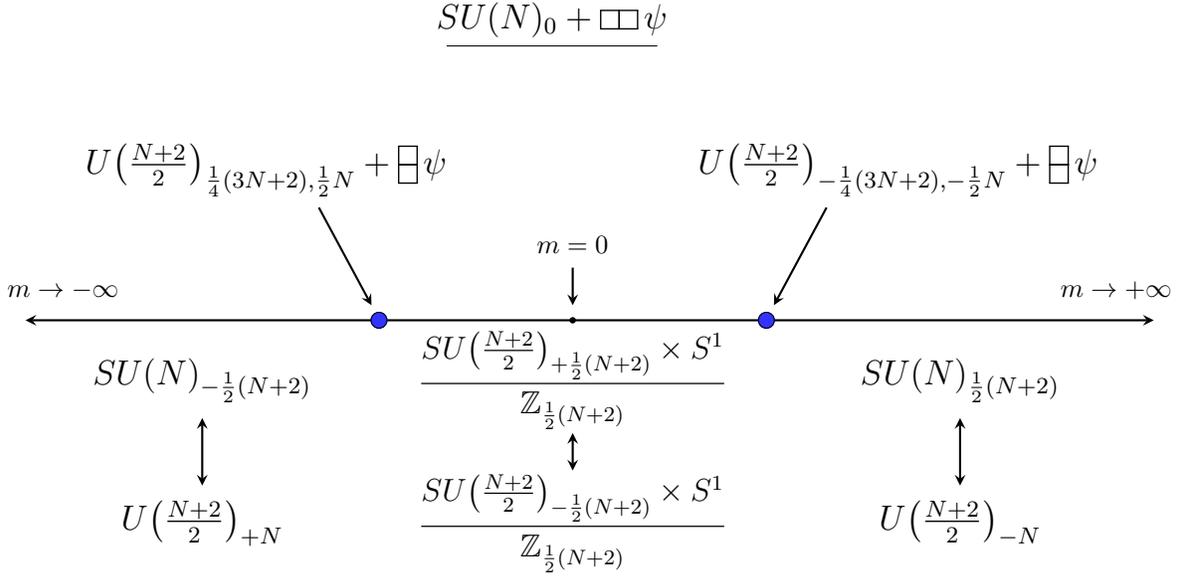

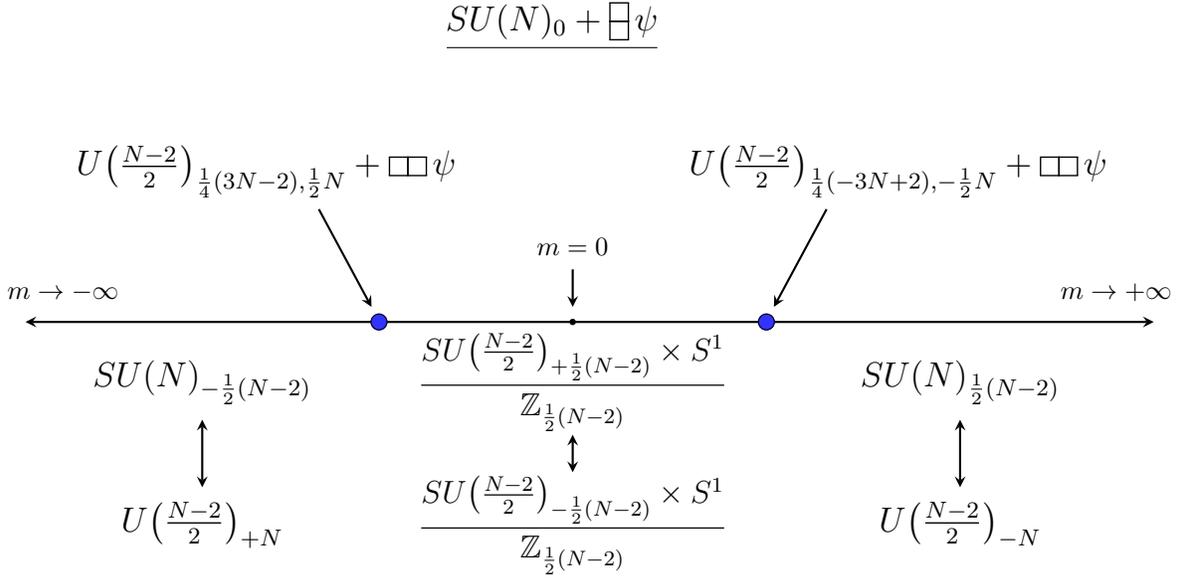
\begin{figure}[!h]
\centering
\begin{tikzpicture}
\node[scale=1.1] at (5.5,4) {$SU(N)_0+\ydiagram{1,1}\,\psi$};
\draw (4.1,3.65) -- (6.9,3.65);
\draw[thick,<->,>=stealth] (-1.5,0) -- (13.5,0);
\filldraw[white!20!blue] (3.2,0) circle (3pt);\draw (3.2,0) circle (3pt);
\filldraw[white!20!blue] (8+.35,0) circle (3pt);\draw (8+.35,0) circle (3pt);
\node[scale=1.1] at (1.7,2) {$U\!\left(\frac{N-2}{2}\right)_{\frac14(3N-2),\frac12N}+\ydiagram{2}\,\psi$};
\node[scale=1.1] at (10.1,2) {$ U\!\left(\frac{N-2}{2}\right)_{\frac14(-3N+2),-\frac12N}+\ydiagram{2}\,\psi$};
\draw[thick,->,>=stealth] (2.4,1.5) -- (3.1,.2);
\draw[thick,->,>=stealth] (8.8+.35,1.5) -- (8.1+.35,.2);

\node[scale=1.1] at (.85,-.8) {$SU(N)_{-\frac12(N-2)}$};
\draw[thick,<->,>=stealth] (.85,-1.3) -- (.85,-2.2);
\node[scale=1.1] at (.85,-2.7) {$ U\!\left(\frac{N-2}{2}\right)_{+N}$};
\node at (5.7+.075,-.8) {$\dfrac{SU\!\left(\frac{N-2}{2}\right)_{+\frac12(N-2)}\times S^1}{\mathbb Z_{\frac12(N-2)}}$};
\draw[thick,<->,>=stealth] (5.7+.075,-1.5) -- (5.7+.075,-2);
\node at (5.7+.075,-2.7) {$\dfrac{SU\!\left(\frac{N-2}{2}\right)_{-\frac12(N-2)}\times S^1}{\mathbb Z_{\frac12(N-2)}}$};
\node[scale=1.1] at (10.925,-.8) {$SU(N)_{\frac12(N-2)}$};
\draw[thick,<->,>=stealth] (10.925,-1.3) -- (10.925,-2.2);
\node[scale=1.1] at (10.925,-2.7) {$ U\!\left(\frac{N-2}{2}\right)_{-N}$};

\node at (5.7+.075,1) {\footnotesize$m=0$};
\draw[thick,->,>=stealth] (5.7+.075,.7) -- (5.7+.075,.2);
\filldraw (5.7+.075,0) circle (1pt);
\node at (13,.4) {\footnotesize$m\to+\infty$};
\node at (-1,.4) {\footnotesize$m\to-\infty$};
\end{tikzpicture}
\caption{Phase diagram of $SU(N)$ gauge theory with a antisymmetric fermion for $k=0$. The circle $S^1$ represents the gapless sigma model with $S^1$ target space. Each phase transition has a dual gauge theory description, which appears with an arrow pointing to the phase transition. The mass deformations are related by 
$m_\psi=-m_{\hat\psi}$ and $m_\psi=-m_{\tilde\psi}$.}
\label{fig:phase_diag_antisymm_zero_k}
\end{figure}

This theory is combined with the Chern-Simons theory in the following way:  the  NGB theory~\eqref{compact} has a non-anomalous $U(1)$ one-form symmetry, corresponding to the conserved two-form current $\epsilon_{\mu\nu\rho}\partial^\rho\phi$. Likewise, 
 $SU(T(R))_{T(R)}$ Chern-Simons theory has a non-anomalous $\mathbb Z_{T(R)}$ one-form symmetry (it is non-anomalous when one views $SU(T(R))_{T(R)}$ as a spin TQFT). We gauge the diagonal $\mathbb Z_{T(R)}$ symmetry, and denote this by
\begin{equation}\label{interzero}
\frac{SU\!\left(T(R)\right)_{T(R)}\times S^1}{\mathbb Z_{T(R)}}\,,
\end{equation}
where $\mathbb Z_{T(R)}$ is the diagonal one-form symmetry. The phase diagrams for $k=0$ are summarized in figures~\ref{fig:phase_diag_symm_zero_k} and~\ref{fig:phase_diag_antisymm_zero_k}.

An immediate consistency check of this scenario -- which we have already discussed in the introduction -- is that the ultraviolet theory is time-reversal symmetric, so it is reassuring to realize that $SU\!\left(\frac N2 \pm 1\right)_{\frac N2 \pm 1}/\mathbb Z_{{\frac N2 \pm 1}}$ Chern-Simons theory and the NGB theory are time-reversal invariant. The time-reversal invariance of the quotient $SU\!\left(\frac N2 \pm 1\right)_{\frac N2 \pm 1}/\mathbb Z_{{\frac N2 \pm 1}}$ can be shown from level/rank duality as in~\cite{Aharony:2016jvv}.

The main feature of the $k=0$ model is, of course, the nonperturbative spontaneous breaking of the $U(1)_B$ baryon number symmetry. The $U(1)_B$ symmetry breaking occurs due to the condensation of a baryon. We will discuss the baryon operators in this theory very briefly in the last section. 

This spontaneous symmetry breaking is not in contradiction with the Vafa-Witten type theorems~\cite{Vafa:1984xg}. In essence, if a symmetry cannot be preserved by a time-reversal invariant mass term then there is no obstruction for the spontaneous breaking of that symmetry. 

Indeed, in our class of theories, it is not possible to deform  by a mass term while preserving both $U(1)_B$ baryon number symmetry and time-reversal symmetry. This is obvious for $N>2$ since there is no mass term whatsoever that preserves time-reversal symmetry. However, the case of $N=2$ requires special attention. 
In the case of $N=2$ with an antisymmetric fermion the theory is always in the large $k$ two-phase regime, and there is no quantum phase and no spontaneous breaking occurs, of course, since the fermion is completely decoupled. 
For $N=2$ with a symmetric fermion, the situation is more interesting. A Dirac fermion in the rank-two symmetric representation is equivalent to {\it two} Majorana fermions in the adjoint representation. Let us denote the two Majorana fermions by $\Psi_1, \Psi_2$, such that $\psi=\Psi_1+i\Psi_2$. The $U(1)_B=SO(2)_B$ baryon symmetry simply rotates these two real fermions 
\begin{equation}\begin{pmatrix}\Psi_1\\ \Psi_2\end{pmatrix}\mapsto\begin{pmatrix}\phantom-\cos\theta & \sin\theta \\ -\sin\theta & \cos
\theta\end{pmatrix}\begin{pmatrix}\Psi_1\\ \Psi_2\end{pmatrix}~.\end{equation}
Time-reversal symmetry can be taken to act as $\Psi_1\mapsto \gamma_0\Psi_1$, and $\Psi_2\mapsto \gamma_0\Psi_2$. Finally, we have a charge-conjugation symmetry $\mathsf C$ that acts as $\Psi_2\mapsto - \Psi_2$ while keeping $\Psi_1$ intact. The minimal scalar baryon operator transforms with charge 2  under $U(1)_B$. The hermitian combinations $i(\bar\Psi_1\Psi_1-\bar\Psi_2\Psi_2)$, and $i\bar\Psi_1\Psi_2$ are the components of this baryon operator.\footnote{They are the real and imaginary parts of the baryon constructed with a Dirac spinor in the symmetric representation $\epsilon_{\alpha_1\alpha_2}\epsilon_{\beta_1\beta_2}\psi^{\alpha_1\beta_1}\psi^{\alpha_2\beta_2}$.} The hermitian combination $i(\bar\Psi_1\Psi_1+\bar\Psi_2\Psi_2)$ is instead invariant under baryon symmetry, but obviously breaks time-reversal symmetry when added to the Lagrangian.

Therefore, clearly, if we want to preserve a time-reversal symmetry, we must use the hermitian  baryon operators above. Indeed, for example, adding $i\bar\Psi_1\Psi_2$ to the Lagrangian would preserve $\mathsf{CT}$. However, there is no way to add a time-reversal invariant mass term that also preserves the $U(1)_B$ baryon symmetry. Therefore, there is no obstruction for baryon symmetry to be spontaneously broken. Interestingly, in the case of $N=2$, the TQFT trivializes (see figure~\ref{fig:phase_diag_symm_zero_k}) because $ SU(2)_{2}/\mathbb Z_2 = SO(3)_1$, which is a trivial spin TQFT. The fact that in the particular case $N=2$ the NGB is not accompanied by a TQFT will be crucial later, when we make contact with $3+1$ dimensional physics. Note also that in the case of $N=2$ it is quite clear that the operator which condenses and leads to the NGB is (without loss of generality) $i(\bar\Psi_1\Psi_1-\bar\Psi_2\Psi_2)$.

\section{Additional Consistency Checks
}\label{sec:additional_consistency_checks}

\subsection{Special Cases}

Here we discuss special values of $N$ where we can compare our proposed dynamics with previously conjectured phase diagrams for other families of theories. We also embed  $SU(2)$ with a rank-two symmetric fermion in a renormalization group flow of ${\cal N}=2$ supersymmetric $SU(2)$ pure gauge theory, and contrast our proposed phase diagram with the expected infrared dynamics of the supersymmetric theory.

\medskip
\noindent
$\bullet$ {\it $SU(4)_k$ with antisymmetric fermion}
\medskip

A  consistency check on our proposed dynamics follows from the isomorphism $SU(4)\simeq Spin(6)$ and the fact that  the antisymmetric representation of $SU(4)$ is the six-dimensional, vector, real representation of $Spin(6)$. Since $\psi$ is a Dirac fermion we have the following equivalence of theories 
\begin{equation}
SU(4)_k+\,\text{antisymmetric}~\psi \equiv Spin(6)_k+\,(N_f=2)~\Psi\,,
\end{equation}
where by $N_f=2$ we mean two Majorana fermions in the vector representation of $Spin(6)$.

 For $k\geq 1$ the phase diagram of $SU(4)_k+\,\text{antisymmetric}$ has two asymptotic phases (see figure~\ref{asymmlargek}). 
The phase diagram of $Spin(6)_k+~N_f~\Psi$ was derived in~\cite{Cordova:2017vab}. For $N_f=2$ and $k\geq 1$ 
the phase diagrams agree trivially by virtue of the identity of the TQFTs  $SU(4)_{n}=Spin(6)_n$. 

We now proceed to the  nontrivial matching for $k=0$ where both theories have an intermediate phase, which we want to compare. Plugging $N=4$ in figure~\ref{asymmsmallk} we find that the intermediate phase of $SU(4)_k+\,\text{antisymmetric}~\psi$ is described by 
\begin{equation}
U(1)_0\,,
\end{equation}
which, as we explained in detail, is simply a free compact scalar $\phi$ with periodicity $2\pi$. 
By contrast, the intermediate phase of $Spin(6)_k+\,N_f~\Psi$ is described by the following coset~\cite{Cordova:2017vab}
\begin{equation}\label{IRSpin}
\frac{SO(N_F)}{S\!\left(O\!\left(\frac{N_F}{2}\right)\times O\!\left(\frac{N_F}{2}\right)\right)}+6\Gamma_{WZ}\,,
\end{equation}
where $\Gamma_{WZ}$ is a Wess-Zumino term. The coset in~\eqref{IRSpin} can be described more explicitly by the equivalence relation of $SO(N_F)$ matrices $O$
\begin{equation}
O\sim P\cdot O\,,
\end{equation}
 where $P$ is a block-diagonal matrix with two $\frac{N_F}{2}\times \frac{N_F}{2}$ blocks $A$ and $B$
 \begin{equation}
 P=\begin{pmatrix}
 A&0\\
 0&B
 \end{pmatrix}
 \end{equation}
such that $\det(A\cdot B)=1$.

For $N_f=2$ the Wess-Zumino term vanishes (because $\pi_3(S^1)=0$) and we are left with a sigma model on the space of $SO(2)$ matrices $O=\begin{pmatrix}\phantom-\cos \theta&\sin\theta\\ -\sin\theta&\cos \theta\end{pmatrix}$ subject to the equivalence relation
\begin{equation}
\begin{pmatrix}\phantom-\cos \theta&\sin\theta\\ -\sin\theta&\cos \theta\end{pmatrix}\sim \begin{pmatrix}-1&0 \\0&-1\end{pmatrix}\cdot \begin{pmatrix}\phantom-\cos \theta&\sin\theta\\ -\sin\theta&\cos \theta\end{pmatrix}\,,
\label{quotient}
\end{equation}
with implies that $\theta\sim \theta+\pi$. Therefore, after the quotient the space is still isomorphic to a circle (albeit with a radius smaller by a factor of 2). Therefore this precisely coincides with the result that we obtained for $SU(4)$ with an antisymmetric fermion. 

In summary, our phase diagram for $SU(N)_k$ with a fermion in the antisymmetric representation for $N=4$ precisely matches the proposed phase diagram of $Spin(M)$ with $N_f$ Majorana fermions in the vector representation for $M=6$ and $N_f=2$. This supports the validity of both phase diagrams.

\medskip
\noindent
$\bullet$ {\it $SU(3)_k$ with an antisymmetric fermion}
\medskip

A somewhat more trivial consistency check can be made for $SU(3)_k$ with an antisymmetric fermion by noting that the rank-two antisymmetric representation of $SU(3)$ is the same as the complex conjugate of the fundamental, three dimensional representation of $SU(3)$. Thus we have the equivalence of theories 
\begin{equation}
SU(3)_k+~\text{antisymmetric}~\psi \equiv SU(3)_k+~(N_f=1)~\Upsilon\,,
\end{equation}
where $\Upsilon$ is a Dirac fermion. These theories, which always have two phases (i.e.~there is no intermediate phase regime), can be seen to have the same phase diagram by comparing figure~\ref{asymmsmallk} for $N=3$ with the phase diagram of $SU(N)_k$ with $N_f$ fermions in the fundamental representation for $N=3$ and $N_f=1$ in~\cite{Komargodski:2017keh}.

\medskip
\noindent
$\bullet$ {\it $SU(2)_k$ with an antisymmetric fermion}
\medskip

A very degenerate special case is $SU(2)_k$ with a fermion in the antisymmetric representation. In this case the 
fermion is decoupled from the gauge dynamics and there is no intermediate phase. The infrared is captured by $SU(2)_k$ Chern-Simons theory except at one point in the phase diagram, which coincides with the phase transition in figure~\ref{asymmlargek}. The phase transition simply corresponds to a neutral fermion becoming massless.

\medskip
\noindent
$\bullet$ {\it $SU(2)_k$ with a symmetric fermion}
\medskip

The rank-two symmetric representation of $SU(2)$ is the adjoint representation. Therefore, figures~\ref{symmlargek} and~\ref{symmsmallk} 
for $N=2$ describe the infrared dynamics of $SU(2)$ QCD with $N_f=2$ adjoint Majorana fermions.\footnote{We recall that our theory is based on a Dirac fermion and therefore $N_f=2$ Majorana adjoint fermions of $SU(2)$.}  This extends the  phase diagram of  $SU/SO/Sp$ adjoint QCD with $N_f=1$ put forward in~\cite{Gomis:2017ixy}. 

The particular case $N=2$ with a symmetric fermion admits an embedding in the ${\mathcal N}=2$ supersymmetric theory  of one $SU(2)$ vector multiplet. In addition to the fields we have in our theory, this model also has a real scalar field $\varphi$ in the adjoint representation, with couples to the fermions via Yukawa terms. What we called $U(1)_B$ is naturally referred to as the $R$-symmetry in the supersymmetric context. We can flow from the supersymmetric theory to our theory by simply  adding a (supersymmetry-breaking) mass term for the real scalar field $\varphi$.\footnote{A closely related ${\cal N}=1$ preserving mass deformation was analyzed in~\cite{Bashmakov:2018wts}.} Below we  analyze what happens if that mass term is very small compared to the scale set by the gauge coupling. 

The infrared of the supersymmetric model consists of~\cite{Affleck:1982as} a complex scalar field $Y$, whose imaginary part transforms inhomogeneously under the $U(1)$ $R$-symmetry, thus signaling spontaneous symmetry breaking of this symmetry. The kinetic term for $Y$ is approximately canonical for large $\re(Y)$ and the potential is the runaway potential $V\sim e^{- \frac{1}{g}\re(Y)}$. Adding a small (supersymmetry-breaking) mass term $m^2\varphi^2$ in the ultraviolet translates to adding a small (supersymmetry-breaking) mass term $m^2\re(Y)^2$ in the infrared. (The map between the deformations in the UV and IR is rather simple for large $\re(Y)$ because the theory is weakly coupled there.) For small enough mass of $\varphi$ the minimum of
\begin{equation}
V=e^{-\frac{1}{g}\re(Y)}+m^2 (\re(Y))^2
\end{equation}
is therefore at large $\re(Y)$ and we can analyze the physics semiclassically. The fermions are all lifted due to the Yukawa couplings and $\re(Y)$ is likewise massive at the minimum of the potential. The deep infrared theory therefore consists of just the (compact) Nambu-Goldstone boson $\im(Y)$ without an additional TQFT, exactly as in the scenario we proposed above for $SU(2)$ with a Dirac fermion in the rank-two symmetric representation.

We shall return to this theory later when we discuss domain walls in $3+1$ dimensional $SU(2)$ Yang-Mills with $N_f=2$ adjoint Majorana fermions.

\subsection{Gravitational Counterterm Matching}\label{sec:gravitational_matching}

Another nontrivial check of our proposed phase diagrams in figures~\ref{symmsmallk} and~\ref{asymmsmallk} can be devised by coupling the theories to background gravity. A well-defined (scheme independent) observable is the difference in the gravitational counterterm  $2\Delta c\, \mathrm{CS_{grav}}$  between the asymptotic negative and asymptotic positive mass phases. This difference is closely related to the difference in the thermal conductance in the two phases. This is an interesting quantity to study because it can be easily computed in the original ``electric variables'' where it is one loop exact. But it can also be computed in the dual variables, followed by traversing the quantum phase, and using the dual variables again. Therefore, one can devise a concrete nontrivial consistency check. Such computations were done in the context of supersymmetric dualities in~\cite{Closset:2012vp,Closset:2012vg}, where the connection to the physically observable thermal conductance (or the analogous charge conductance which we will study soon) is explained.

The jump in the gravitational counterterm in the electric variables is given by the number of fermions in the ultraviolet gauge theory (this is related to the parity anomaly~\cite{Niemi:1983rq,Redlich:1983kn,Redlich:1983dv}). Therefore, in our theories
 \begin{equation}
 \Delta c=-\dim(R)~,
 \label{directj}
 \end{equation}
where $R$ is either the rank-two symmetric or antisymmetric representation, respectively.

Our phase diagrams for $k< T(R)$ in figures~\ref{symmsmallk} and~\ref{asymmsmallk} provide us with another way to compute this difference. The two computations must  agree  for consistency. We start with the TQFT in the asymptotic negative mass phase $SU(N)_{k-T(R)}$ and work our way towards the asymptotic positive mass phase $SU(N)_{k+T(R)}$. This requires tracking the jump of the gravitational counterterm across level/rank dualities, where a gravitational counterterm is generated, and across the phase transitions:

\begin{itemize}
\item $SU/U$ TQFT level/rank duality in the asymptotic negative phase~\cite{Hsin:2016blu}: 
\begin{equation}
\Delta c_1= N(k-T(R))
\end{equation}
\item Jump induced from positive to negative mass of the leftmost  dual gauge theory: 
\begin{equation}
\Delta c_2= \dim(\hat R)
\end{equation}
\item $U/U$ TQFT level/rank duality in the intermediate phase~\cite{Hsin:2016blu}: 
\begin{equation}
\Delta c_3= (T(R)-k)(T(R)+k)-1
\end{equation}
\item Jump induced from positive to negative mass of the rightmost dual gauge theory: 
\begin{equation}
\Delta c_4= \dim(\tilde R)
\end{equation} 
\item $U/SU$ TQFT level/rank duality in the asymptotic positive phase~\cite{Hsin:2016blu}: 
\begin{equation} 
\Delta c_5=-(k+T(R))N
\end{equation}
\end{itemize}
Here $\hat R$ and $\tilde R$ denote the representation of the fermion in the leftmost and rightmost dual descriptions in figures~\ref{symmsmallk} and~\ref{asymmsmallk}.

Adding up the contributions to the jump following this path we find that it precisely matches that in~\eqref{directj}
\begin{equation}\label{matchingg}
\sum_{I=1} ^5 \Delta c_I\equiv \Delta c\,,
\end{equation}
both for the theory with a symmetric and antisymmetric fermion! 

\subsection{Baryon Counterterm Matching}

An entirely analogous exercise to~\eqref{matchingg} is to match the difference in the baryon number conductivity coefficient. This coefficient is simply the difference between the two asymptotic phases in the Chern-Simons term for the baryon background gauge field $B$, i.e., we are after the difference
\begin{equation}
\frac{\Delta \kappa}{4\pi}\int B\wedge \mathrm dB\,.
\end{equation}
Here we think about $B$ as an ordinary $U(1)$ connection and the space-time is assumed to be a spin manifold.\footnote{As all our theories are fermionic and the baryon number clearly satisfies a spin-charge relation, we can in principle also study our theories on spin$_c$ manifolds. This leads to some additional nontrivial consistency checks which we do not present here.}

We need to carefully normalize the baryon charge of the fermion. The most convenient choice is to imagine that the fermion is in the (anti-)symmetric representation of $U(N)$ rather than $SU(N)$ and the diagonal of $U(N)$ is the baryon number. This would lead to the fermion carrying charge $2$. However, the corresponding baryon gauge field would then have possible fractional fluxes and in order to fix that we need to take the charge to be $2/N$.

We can therefore compute $\Delta \kappa$ straightforwardly in the electric variables as
\begin{equation}\label{eq:delta_kappa}
\Delta \kappa=\left(\frac{2}{N}\right)^2\dim (R)=\frac{4}{N^2}\frac12 N(N\pm 1)=\frac{2}{N}(N\pm 1) \,.
\end{equation}

As before, we can also compute $\Delta\kappa$ using the dual description:
\begin{itemize}
\item First, in the phase with large negative mass we need to perform level/rank duality between $SU(N)_{k-T(R)}$ and $U(T(R)-k)_N$, which leads to a jump\footnote{Let us explain briefly how to derive this shift from~\cite{Hsin:2016blu}. Using the notation of~\cite{Hsin:2016blu}, the Lagrangian for $SU(N)_K$ is
\begin{equation}
\mathcal L_{SU(N)_K}=\frac{K}{4\pi}\Tr\left[b\mathrm db-\frac23ib^3\right]+\frac{\epsilon_K}{4\pi}(\Tr b)\mathrm d(\Tr b)+\frac{1}{2\pi}c\mathrm d(\Tr b+B)
\end{equation}
where $b$ us a $\mathfrak u(N)$ gauge field. If we integrate out $c$ and remove the trace $b:=\tilde b-\frac{1}{N} B$ (with $\Tr\tilde b\equiv0$), we get
\begin{equation}
\mathcal L_{SU(N)_K}\to\frac{K}{4\pi}\Tr\left[\tilde b\mathrm d\tilde b-\frac23i\tilde b^3\right]+\frac{K}{4\pi N}B\mathrm dB+\frac{\epsilon_K}{4\pi}B\mathrm dB\,.
\end{equation}
The level/rank dual $U(K)_{-N}$ also has a term $\frac{\epsilon_K}{4\pi}B\mathrm dB$, so the relative shift by the contact term is only given by the term $\frac{K/N}{4\pi}B\mathrm dB$.}~\cite{Hsin:2016blu}
\begin{equation}
\Delta \kappa_1=-\frac{k-T(R)}{N}=-\frac{k-\frac12({N\pm 2})}{N}\,.
\end{equation}

\item Next, there is a crucial difference with our computation of the thermal conductivity. Since the baryon symmetry maps to the magnetic symmetry in the dual variables, and since the dual fermions are not charged under the magnetic symmetry, we get that $\Delta \kappa_2\equiv\Delta\kappa_4\equiv0$. In other words, the phase transitions do not lead to a jump in the conductivity.

\item Next we need to address $\Delta \kappa_3$, which arises from the $U/U$ level/rank duality in the quantum phase. We find again from~\cite{Hsin:2016blu} that
\begin{equation}
\Delta \kappa_3=1\,.
\end{equation}

\item Finally, the positive-mass $SU/U$ level/rank duality leads to
\begin{equation}
\Delta \kappa_5= \frac{k+\frac12({N\pm 2})}{N}\,.
\end{equation}

\end{itemize}

We see that if we add all the partial jumps $\Delta\kappa_I$ we get precisely the same shift in the baryon counterterm~\eqref{eq:delta_kappa} computed in the electric theory:
\begin{equation}
\sum_{I=1}^5\Delta\kappa_I\equiv\Delta\kappa
\end{equation}
for both the symmetric and antisymmetric representation!

The matching of the gravitational contact term guarantees that the phase diagram remains consistent in curved space and the thermal conductivities are single-valued, as they should be in physical theories. 
The matching of the baryon contact term further guarantees that we can consistently gauge the ultraviolet baryon symmetry in all the phases. Therefore, one can derive from our phase diagrams and dualities also the phase diagrams and corresponding dualities for $U(N)$ gauge theories coupled to two-index matter fields.

\section{Domain Walls in Four Dimensional Gauge Theories}

Let us consider the four-dimensional theory of a Dirac fermion in the symmetric/antisymmetric representation coupled to $SU(N)$ gauge fields. We can equivalently think about it as $SU(N)$ gauge theory coupled to a Weyl fermion in the symmetric/antisymmetric representation and another Weyl fermion in the conjugate representation. Let us begin with the massless theory. This theory has a ($\mathbb Z_{2(N-2)}$) $\mathbb Z_{2(N+2)}$ discrete chiral symmetry that acts by re-phasing the two Weyl fermions together, 
\begin{equation}
\mathbb Z_{2(N\pm2)}:\ \psi,\tilde \psi\, \mapsto\, e^{i\alpha} \psi, e^{i\alpha} \tilde \psi ~,\qquad \alpha = \frac{2\pi k}{2(N\pm 2)}~,\ k\in \mathbb Z~.
\end{equation}
In addition the theory enjoys baryon number symmetry $U(1)_B$ which acts by re-phasing the two fermions in an opposite fashion
\begin{equation}
U(1)_B:\ \psi,\tilde \psi\, \mapsto\, e^{i\beta} \psi, e^{-i\beta} \tilde \psi ~.
\end{equation}

The special case of $N=2$ with a symmetric representation Dirac fermion is equivalent to $SU(2)$ gauge theory with two Weyl fermions in the adjoint representation. In this case the $U(1)_B$ symmetry is in fact enhanced to $SU(2)_F$ flavor symmetry (and in addition, there is the discrete $\mathbb Z_8$ axial symmetry, where the order-two generator in $\mathbb Z_8$ is identified with the center of $SU(2)_F$ flavor symmetry. This order-two generator coincides with the fermion number symmetry, and it is hence unbreakable as long as the vacuum is Poincar\'e invariant.)

These theories admit a mass deformation, $M\psi\tilde \psi$, and we can take $M$ to be non-negative without loss of generality, at the expense of having to keep track of the $\theta$ parameter of the gauge theory. The mass perturbation breaks the $\mathbb Z_{2(N\pm2)}$ symmetry down to $\mathbb Z_2$. However, the mass term preserves $U(1)_B$. For $\theta=0,\pi$ also the time-reversal symmetry is preserved.

In the special case of $N=2$, the mass perturbation breaks $SU(2)_F$ symmetry, but it preserves baryon number, which can be identified with the Cartan subgroup of $SU(2)_F$. 
The Vafa-Witten-like theorems would imply that the massless theory cannot break $U(1)_B$. We will assume that $SU(2)_F$ is broken to $U(1)_B$. 

It is reasonable to assume that the massless theory  breaks the chiral symmetry $\mathbb Z_{2(N\pm 2)}$ as\footnote{This can be proven in the planar limit~\cite{Armoni:2003fb}, where the theory in the meson sector is equivalent to $\mathcal{N}=1$ Supersymmetric Yang Mills theory, which is known to develop a condensate. Therefore our statement about the symmetry breaking pattern certainly holds for large enough finite $N$.} 
\begin{equation}
\mathbb Z_{2(N\pm 2)}\to \mathbb Z_2\,. 
\end{equation}

According to these assumptions, the vacuum structure of the theory is therefore:

\begin{itemize}
\item $N>2$: The massless theory has $N\pm 2$ vacua, each of which is trivial and gapped. The order parameter distinguishing these vacua is the fermion bilinear $\langle \psi\tilde \psi\rangle$, which is charged under $\mathbb Z_{2(N\pm 2)}/ \mathbb Z_2 $.
\item $N=2$: Here $SU(2)_F$ breaks spontaneously to $U(1)_B$ but also the axial symmetry is spontaneously broken. The fermion bilinear is in the adjoint representation and it is the order parameter leading to this symmetry breaking pattern.\footnote{Indeed, since the order parameter must be a scalar in space-time the Lorentz indices are contracted antisymmetrically, and since it must be gauge invariant, the gauge indices are contracted symmetrically and hence the flavor indices must be contracted symmetrically as well, leading to the symmetric product of the fundamental representation of $SU(2)_F$ with itself, namely, the adjoint representation.} Let us parameterize it without loss of generality by $\langle \psi\psi\rangle=\left(\begin{matrix}1 & 0 \\ 0 & -1 \end{matrix}\right)$.
This leads to the breaking pattern $SU(2)_F\to U(1)_B$, and the corresponding coset manifold is $S^2$. Acting with the generator of $\mathbb{Z}_8$ we get a new vacuum, and therefore we have at least two copies of the coset $S^2$. However, acting with the square of the generator of $\mathbb{Z}_8$ we get the matrix $\left(\begin{matrix}-1 & 0 \\ 0 & 1 \end{matrix}\right)$, but this is in fact on the same coset as the original condensate (more precisely, the Weyl group of $SU(2)_F$ relates these two configurations). Hence we have exactly two copies of $S^2$, and the broken axial symmetry allows us to move from one copy to the other copy.\footnote{This scenario of the $SU(2)$ gauge theory with two adjoint fermions flowing in the infrared to two copies of $S^2$ has been recently connected to the Seiberg-Witten solution of the $\mathcal{N}=2$ vector multiplet theory~\cite{Cordova:2018acb}. Other possibilities for the infrared dynamics were recently discussed also in~\cite{Bi:2018xvr, Anber:2018tcj}. }

\end{itemize}

Let us now turn on a small positive mass $M$. This corresponds to a small potential on the above space of vacua which is a function of $M,\theta$. For $N>2$, for any $\theta\neq 0$ this lifts the degeneracy and picks up one of the $N\pm 2$ vacua. At $\theta=\pi$ there is a first order transition and two (adjacent) vacua are exactly degenerate. While this analysis is done at small $M$, it is natural to assume that this is true for any positive $M$. In particular, at asymptotically large $M$ this agrees with the expectations from pure Yang-Mills theory, which is supposed to have a trivial ground state for any $\theta\neq \pi$ and two degenerate vacua at $\theta=\pi$. 

For $N=2$ we can again turn on some mass $M$ and fix the theta angle. But now the mass $M$ is an adjoint $SU(2)$ matrix. Without loss of generality we take this matrix to be in the Cartan and hence the eigenvalues are real (and we keep track of $\theta$). Therefore we choose $M=M_0\left(\begin{matrix}1 & 0 \\ 0 & -1 \end{matrix}\right)$ with positive $M_0$. The physics depends only on the combination
\begin{equation}
M_0e^{i\theta/4}
\end{equation}
and, furthermore, all the physical observables must be periodic under $\theta\to \theta+2\pi$.

We can parameterize the vacuum configurations by the adjoint $SU(2)_F$ matrix of fermion condensates 
\begin{equation}
\mathcal M_\mathrm{VAC}=\biggl\{U\left(\begin{matrix}1 & 0 \\ 0 & -1 \end{matrix}\right)U^{-1}\biggr\}\cup \biggl\{V\left(\begin{matrix}i & 0 \\ 0 & -i \end{matrix}\right)V^{-1}\biggr\}
\end{equation}
where $U,V$ are $SU(2)$ matrices. This is just the union of two $S^2$'s, as we explained above. The potential (up to an unimportant proportionality factor) induced by the deformation by $M,\theta$ on the space $\mathcal M_\mathrm{VAC}$ is 
\begin{equation}
E_1=M_0e^{i\theta/4}\Tr\left(\left(\begin{matrix}1 & 0 \\ 0 & -1 \end{matrix}\right) U\left(\begin{matrix}1 & 0 \\ 0 & -1 \end{matrix}\right)U^{-1}\right)+c.c.
\end{equation}
on the first $S^2$ and 
\begin{equation}
E_2=iM_0e^{i\theta/4}\Tr\left(\left(\begin{matrix}1 & 0 \\ 0 & -1 \end{matrix}\right) V\left(\begin{matrix}1 & 0 \\ 0 & -1 \end{matrix}\right)V^{-1}\right)+c.c.
\end{equation}
on the second $S^2$. We need to minimize over $U$ and $V$ and then find the global minimum by comparing these two sectors. First we simplify the expressions for $E_{1,2}$ and write (again, up to unimportant proportionality factors)
\begin{equation}
\begin{aligned}
E_1&=M_0 \cos(\theta/4)\Tr\left(\left(\begin{matrix}1 & 0 \\ 0 & -1 \end{matrix}\right) U\left(\begin{matrix}1 & 0 \\ 0 & -1 \end{matrix}\right)U^{-1}\right)\,,\\
E_2&=M_0 \sin(\theta/4)\Tr\left(\left(\begin{matrix}1 & 0 \\ 0 & -1 \end{matrix}\right) V\left(\begin{matrix}1 & 0 \\ 0 & -1 \end{matrix}\right)V^{-1}\right)\,.
\end{aligned}
\end{equation}

Using the Cauchy-Schwarz inequality we see that the expression in the trace is minimized when the gaugino condensate is $\langle \psi\psi\rangle=\left(\begin{matrix}-1 & 0 \\ 0 & 1 \end{matrix}\right)$ and it is maximized when the gaugino condensate is $\langle \psi\psi\rangle=\left(\begin{matrix}1 & 0 \\ 0 & -1 \end{matrix}\right)$. We refer to these two points on $S^2$ as the south and north pole of $S^2$.
We therefore have the following phases for small enough $M_0$, as a function of $\theta$:
\begin{itemize}
\item $-\pi<\theta<\pi$: The true vacuum is at the south pole of the first $S^2$. 
\item $\pi<\theta<3\pi$: The true vacuum is at the south pole of the second $S^2$
\item $3\pi<\theta<5\pi$: The true vacuum is at the north pole of the first $S^2$
\item $5\pi<\theta<7\pi$: The true vacuum is at the north pole of the second $S^2$
\item $\theta=\pi,3\pi,5\pi,7\pi$: The two vacua on the two sides of the transition are exactly degenerate.
\end{itemize}
While the periodicity of the above list of vacua is $8\pi$, of course the physical observables are $2\pi$ periodic.
In addition, as before, while our analysis is reliable for small $M_0$, it is consistent to assume that the bulk vacua behave as above for any $M_0$. 

We now turn to analyzing domain walls in this theory. We can start from the massless case, which is the most difficult (and the richest) case. We consider first $N>2$. Since we have $N\pm2$ vacua, we can study the domain wall between any pair of vacua. However, using the spontaneously broken axial symmetry we see that the result only depends on the difference of the phases of the gaugino condensates on the two sides. Let us label the vacua by $J=1,...,N\pm2$ according to the phase of the guagino condensate $\langle \psi\tilde \psi\rangle= e^{\frac{2\pi i J}{N\pm2}} $.
A natural conjecture for the domain wall theory is to identify it with the quantum phase in the corresponding $2+1$ dimensional gauge theory of $SU(N)$ with a symmetric (antisymmetric) fermion. 
This leads to the proposal that the theory on the domain wall connecting the vacuum $I$ and the vacuum $J$. We can choose the orientation such that without loss of generality, say, $I>J$ and $J$ is on the right hand side of the wall. The other cases are obtained by simply reversing the orientation. Then the domain wall theory is 
\begin{equation}\label{AVSAS} U\left(I-J\right)_{N\pm 2 -I+J, N\pm 2 -2I+2J}~.\end{equation}
Of course, this TQFT has to be accompanied by the decoupled center of mass degree of freedom (which is described to leading order by the Nambu-Goto action). 

The proposal~\eqref{AVSAS} is a generalization of the Acharya-Vafa theory for SYM~\cite{Acharya:2001dz}. Note a few interesting facts that follow from~\eqref{AVSAS}.

\begin{itemize}

\item If we take $I=J+1$ we obtain the TQFT $U(1)_{N}$ in the case of the symmetric fermion, and $U(1)_{N-4}$ in the case of the antisymmetric fermion. Note that for finite (positive) $M$, and $\theta=\pi$ precisely these two adjacent vacua are degenerate and hence this domain wall continues to exist also at finite $M$. At very large $M$ we can integrate out the fermion and remain with the pure Yang-Mills theory, where the domain wall is given by $U(1)_N$ (more precisely, the domain wall of pure Yang-Mills theory is level/rank dual to the $U(1)_N$ TQFT). We see that in the symmetric fermion case no phase transition occurs. This is similar to the theory with the adjoint fermion. But in the case of the antisymmetric fermion we see that a phase transition does occur as we crank up the mass. If the phase transition is second order, it would be natural to assume that it is in the same universality class as the corresponding 3d phase transition, namely, it is given by $U(1)_{N-2}$ plus a charge-$2$ fermion. This discussion (beautifully) makes sense also for the degenerate case $N=2$, where the corresponding theory with the antisymmetric tensor is equivalent to the pure Yang-Mills theory with a neutral Dirac fermion. The domain wall theory is always $U(1)_2$ (here we use that $U(1)_2\simeq U(1)_{-2}$) and the $U(1)_0$ plus a charge-$2$ fermion leads to a massless fermion on the wall~\cite{Son:2015xqa,Karch:2016sxi,Seiberg:2016gmd,Murugan:2016zal}, which can be thought of as arising precisely from the massless Dirac fermion in the bulk! 

\item In general, combining the four-dimensional spontaneously broken axial symmetry and time-reversal symmetry, we can derive that the domain wall theory connecting $I$ and $J$ should be isomorphic to the one connecting $N\pm2 - I+J$ and $0$. Indeed, this is merely the statement that 
\begin{equation}
U\left(I-J\right)_{N\pm 2 -I+J, N\pm 2 -2I+2J}\simeq U\left(N\pm2-I+J\right)_{J-I, N\pm 2 -2I+2J}
\end{equation}
which is nothing but level/rank duality in three dimensions. 

\item If $N$ is even then there exists a time-reversal invariant domain wall, given by $I-J=\frac{N\pm 2}{2}$. The corresponding theory on the wall is $U\!\left(\frac{N\pm 2}{2}\right)_{\frac{N\pm 2}{2},0}$. We discussed in detail that this theory should be interpreted as a massless NGB coupled to some TQFT. While baryon symmetry is not broken in the bulk, we see that it is broken on the wall! In principle, this domain wall theory is a nonperturbative object and it is hard to understand where this massless NGB comes from and why (recall that in the case of the Acharya-Vafa domain walls, there is no such massless field). We can think of this domain wall theory as a bound state of $\frac{N\pm 2}{2}$ elementary domain walls (the elementary domain walls are the ones described in the first bullet point). There is however one case, $N=2$ with a symmetric fermion, where this massless NGB can be seen explicitly. We describe this mechanism below.

\end{itemize}

Now let us discuss in detail the case of $N=2$ with a Dirac fermion in the adjoint representation. Recall that, as explained above, instead of having $N+2 = 4$ isolated trivial vacua we have two copies of $S^2$. The ``elementary'' domain wall corresponds to connecting the south pole of the first $S^2$ and the south pole of the second $S^2$ (or any isomorphic configuration thereof). This wall is hard to understand since it passes in regions of field space which are not within our effective theory. 
However, our prediction above for the physics of this domain wall is the $U(1)_2$ TQFT, which makes a lot of sense (in the limit of softly deformed pure $\mathcal{N}=2$  SYM theory, this result can be derived directly from the Seiberg-Witten solution along the lines of~\cite{Kaplunovsky:1998vt}).

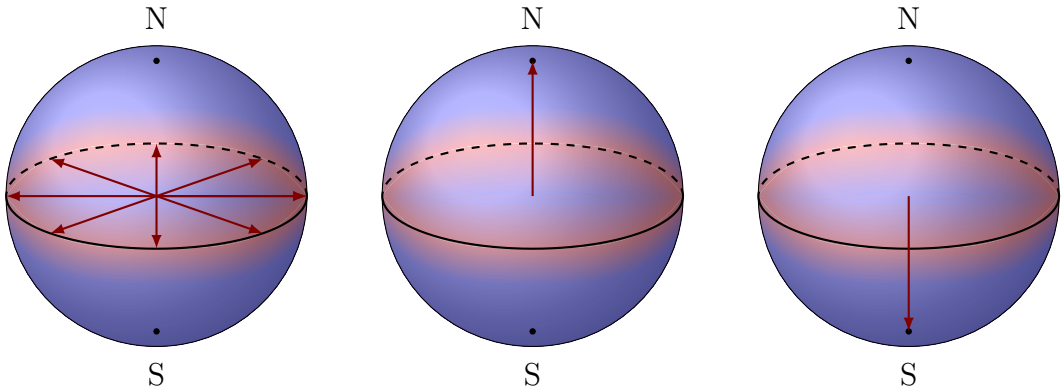
\begin{figure}[!h]
\centering
\begin{tikzpicture}

\foreach \XX in {0,5,10} {
\begin{scope}[shift={(\XX,0)}]
\begin{scope}
\clip (-2,0) to[out=-90,in=180] (0,-2) to[out=0,in=-90] (2,0) to[out=-90-35,in=0](0,-.7) to[out=180,in=-55] (-2,0);
\foreach \x [evaluate=\x as \xn using {\x*10}] in {0,...,10} {
\fill[blue!\xn!red!80!white,yshift=-1.3*\x] (-2,0) to[out=-90,in=180] (0,-2) to[out=0,in=-90] (2,0) to[out=-90-35,in=0](0,-.7) to[out=180,in=-55] (-2,0);
}
\end{scope}
\begin{scope}
\clip (-2,0) to[out=90,in=180] (0,2) to[out=0,in=90] (2,0) to[out=90+35,in=0] (0,.7) to[out=180,in=55] (-2,0);
\foreach \x [evaluate=\x as \xn using {\x*10}] in {0,...,10} {
\fill[blue!\xn!red!80!white,yshift=1.3*\x] (-2,0) to[out=90,in=180] (0,2) to[out=0,in=90] (2,0) to[out=90+35,in=0] (0,.7) to[out=180,in=55] (-2,0);
}
\end{scope}
\begin{scope}
\clip (-2,0) to[out=0,in=180] (0,0) to[out=0,in=180] (2,0) to[out=-90-35,in=0] (0,-.7) to[out=180,in=-55] (-2,0);
\foreach \x [evaluate=\x as \xn using {\x*10}] in {0,...,10} {
\fill[blue!\xn!red!80!white,yshift=2*\x] (-2,0) to[out=90,in=180] (-2,0) to[out=0,in=180] (0,0) to[out=0,in=180] (2,0) to[out=-90-35,in=0] (0,-.7) to[out=180,in=-55] (-2,0);
}
\end{scope}
\begin{scope}
\clip (-2,0) to[out=0,in=180] (0,0) to[out=180,in=180] (2,0) to[out=90+35,in=0] (0,.7) to[out=180,in=55] (-2,0);
\foreach \x [evaluate=\x as \xn using {\x*10}] in {0,...,10} {
\fill[blue!\xn!red!80!white,yshift=-2*\x] (-2,0) to[out=0,in=180] (0,0) to[out=180,in=180] (2,0) to[out=90+35,in=0] (0,.7) to[out=180,in=55] (-2,0);
}
\end{scope}
\shade[ball color=blue!10!white,opacity=.7] (0,0) circle (2cm);
\draw[domain=0:180,smooth,black,dashed,thick]  plot ({2*cos(\x)},{.7*sin(\x)});
\filldraw (0,1.8) circle (1pt) node[above=8pt] {$\mathrm{N}$};
\filldraw (0,-1.8) circle (1pt) node[below=8pt] {$\mathrm{S}$};
\draw (0,0) circle (2cm);
\draw[domain=180:360,smooth,black,thick]  plot ({2*cos(\x)},{.7*sin(\x)});
\end{scope}
}

\foreach \x in {0,45,...,360}
\draw[thick,->,>=latex,red!50!black] (0,0) -- ({2*cos(\x)},{.7*sin(\x)});

\draw[thick,->,>=latex,red!50!black] (5,0) -- (5,1.8);
\draw[thick,->,>=latex,red!50!black] (10,0) -- (10,-1.8);

\end{tikzpicture}
\caption{These figures represent the NG boson that arises on the domain wall connecting the vacuum at the north pole with the vacuum at the south pole. The first figure is the NG boson on the domain wall and the second and third figures represent the bulk vacua.}
\label{azimuthal}
\end{figure}

\begin{figure}[!h]
\centering
\begin{tikzpicture}[scale=2.5]
\shade[shading=ball,opacity=.5,rotate=45,outer color=black!50!blue!60!white,inner color=blue!10!white] (1.414,0) ellipse (1.414cm and .3cm);
\draw[rotate=+45,gray,opacity=.5] (1.414,0) ellipse (1.414cm and .3cm);
\draw[rotate=+45,gray] (0,0) arc[start angle=180,end angle=360, x radius=1.414cm, y radius=.4*.3cm];
\draw[rotate=+45,gray,dotted] (2*1.414,0) arc[start angle=0,end angle=180, x radius=1.414cm, y radius=.4*.3cm];
\fill[white,rotate=-45] (0,1.414) ellipse (1.4*1.414cm and 1.4*.3cm);
\filldraw (0,0) circle (.6pt);
\filldraw (2,0) circle (.6pt);
\filldraw (2,2) circle (.6pt);
\filldraw (0,2) circle (.6pt);
\draw[->,thick,>=latex] (0,0) -- (1.1,0);
\draw[->,thick,>=latex] (2,0) -- (2,1.1);
\draw[->,thick,>=latex] (2,2) -- (.9,2);
\draw[->,thick,>=latex] (0,2) -- (0,.9);
\draw[thick] (1,0) -- (2,0);
\shade[shading=ball,opacity=.5,rotate=-45,outer color=black!50!blue!60!white,inner color=blue!10!white] (0,1.414) ellipse (1.414cm and .3cm);
\draw[rotate=-45,gray,opacity=.5] (0,1.414) ellipse (1.414cm and .3cm);
\draw[rotate=+45,gray,dotted] (1.414+.3,0) arc[start angle=0,end angle=180, x radius=.21*1.414cm, y radius=.2*.3cm];
\draw[rotate=+45,gray] (1.414-.3,0) arc[start angle=180,end angle=360, x radius=.21*1.414cm, y radius=.2*.3cm];
\draw[rotate=-45,gray] (-1.414,1.414) arc[start angle=180,end angle=360, x radius=1.414cm, y radius=.4*.3cm];
\draw[rotate=-45,gray,dotted] (1.414,1.414) arc[start angle=0,end angle=180, x radius=1.414cm, y radius=.4*.3cm];
\draw[thick] (2,1) -- (2,2);
\draw[thick] (1,2) -- (0,2);
\draw[thick] (0,1) -- (0,0);
\end{tikzpicture}
\caption{Moduli space of $SU(2)$ plus a symmetric fermion, consisting of two  copies of $S^2$ (the $S^2$'s in the figure are stretched only for the convenience of the picture). }
\label{boundwall}
\end{figure}
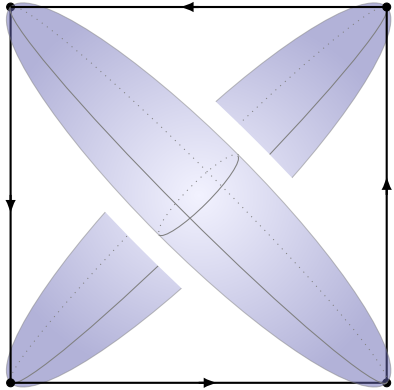

The bound state of two such elementary walls corresponds to a jump of $\theta$ by $4\pi$ and what it does is to connect the north pole and the south pole of the same given $S^2$. Now, the wall can be analyzed entirely within classical field theory, since the wall is merely a geodesic running from the north to the south pole of the $O(3)$ NLSM. The massless bosonic mode that our 3d model predicts is simply the azimuthal degree of freedom of that geodesic trajectory, see figure~\ref{azimuthal}. Therefore, the domain wall has indeed a compact NGB which corresponds to the spontaneous breaking of baryon symmetry. In figure~\ref{boundwall} we depict the four vacua at the north and south poles of the two $S^2$'s as the vertices of the square and we draw the two $S^2$'s that stretch along the diagonals, parameterizing the vacua of the theory. 

For this story to hold up it is essential that the TQFT that accompanies the NGB is trivial. Indeed, it is $PSU(2)_2 = SO(3)_1$ which is a trivial theory. In the four dimensional theory there is nowhere for the domain wall to obtain a Chern-Simons term from since the sphere has the standard non-singular round metric and the domain wall can be understood entirely within effective field theory. Therefore the NGB in the infrared is not accompanied by a TQFT, in accordance with our prediction from the 3d analysis.

Note that the emergence of the NGB on the wall here is quite analogous to the way the symmetry breaking phases of QCD$_3$ emerge from the corresponding four-dimensional construction~\cite{Gaiotto:2017tne}. For similar constructions see also~\cite{Gaiotto:2017yup,Komargodski:2017dmc,Komargodski:2017smk, Draper:2018mpj,Ritz:2018mce, Argurio:2018uup,Anber:2018jdf,,Armoni:2018ahv}

\section{Comments on Future Directions}

Our theories with $m=k=0$ all have a time-reversal symmetry, satisfying 
\begin{equation}
\mathsf T^2=(-1)^F\,.
\end{equation}
It is therefore possible to study these theories on $\mathrm{pin}_+$ manifolds. It is well known that there is a purely gravitational $\mathbb Z_{16}$ time-reversal anomaly in such cases~\cite{Fidkowski:2013jua,Metlitski:2014xqa,Wang:2014lca,KitaevZ16,Witten:2015aba}. We denote the anomaly by $\nu\in \mathbb Z_{16}$. Here we would like to briefly mention some new aspects of this anomaly, which will be explained in more detail elsewhere. 

A naive approach to computing the anomaly $\nu$ in the ultraviolet is to disregard the gauge interactions and compute the total (weighted) number of Majorana fermions. This seems physically justified because the gauge interactions are arbitrarily weak in the deep ultraviolet. However, in~\cite{Gomis:2017ixy} it was noted that this procedure is often ambiguous; it is not invariant under gauge transformations, in the sense that if we compose $\mathsf T$ with a gauge transformation then $\nu$ may change by an integer which is not a multiple of $16$.

In~\cite{Cordova:2017kue} it was pointed out that in theories where the gauge group is not simply connected, the time-reversal symmetry algebra could be deformed by the magnetic symmetry and as a result $\nu$ would not be well defined. Here we will demonstrate that $\nu\in \mathbb{Z}_{16}$ cannot be canonically defined
(in general)  in theories with a one-form symmetry. This even happens in theories without a magnetic symmetry (i.e. based on simply connected gauge groups). The lack of a canonical choice of $\nu$ has to do with the fact that the full symmetry group of the theory has more generators than just ${\mathsf T}$, and there is no canonical way to set to zero the background fields for the one-form symmetry. See below.

Let us consider a simple example, which was already studied in~\cite{Gomis:2017ixy}.   Consider $SU(2)$ gauge theory coupled to a single Majorana adjoint fermion (and $k=m=0$). Let the time-reversal symmetry act in the standard way, which leads to $\nu=3$, simply because there are three Majorana fermions in total.  However, let us now compose $\mathsf T$ with the gauge transformation 
$U=\left(\begin{matrix}0 & 1 \\ -1 & 0\end{matrix}\right)$. The (traceless) Hermitian matrix of Majorana fermions transforms under this gauge transformation as 
\begin{equation}
\left(\begin{matrix}\Psi & \chi \\ \chi^\dagger & -\Psi\end{matrix}\right)\mapsto\left(\begin{matrix}0 & 1 \\ -1 & 0\end{matrix}\right)\left(\begin{matrix}\Psi & \chi \\ \chi^\dagger & -\Psi\end{matrix}\right)\left(\begin{matrix}0 & -1 \\ 1 & 0\end{matrix}\right)=\left(\begin{matrix}-\Psi & -\chi^\dagger \\ -\chi & \Psi\end{matrix}\right)~. 
\end{equation}

If our original time-reversal symmetry acts as $\mathsf T\Psi=\gamma^0\Psi$ and $\mathsf T\chi=\gamma^0\chi^\dagger$ as usual, then combining this with the gauge transformation we see that  $\mathsf T\circ U \Psi = -\gamma^0\Psi$ and $\mathsf T\circ U \chi = -\gamma^0\chi$.
Remembering that $\mathsf T\circ U$ is antilinear (squaring to $(-1)^F$) we thus see that now it would appear that the time-reversal anomaly is $\nu = -1$.
We therefore clearly see that in theories with a one-form symmetry the $\nu\in \mathbb Z_{16}$ anomaly is not uniquely defined in spite of the fact that there is no magnetic symmetry. The same phenomenon takes place in many of the examples studied in~\cite{Gomis:2017ixy,Cordova:2017kue}. We will now give a physical as well as a mathematical interpretation of this  phenomenon. 

The $SU(2)$ gauge theory coupled to the adjoint Majorana fermion was claimed to flow in the infrared to a free massless Majorana fermion alongside with the $U(1)_2$ TQFT~\cite{Gomis:2017ixy}. The time-reversal anomaly of the $U(1)_2$ TQFT could be either $\nu=+2$ or $\nu=-2$ -- this depends on the time-reversal transformation of the semion in the TQFT.  The contribution of the decoupled Majorana fermion is always $\nu_\mathrm{Majorana}=+1$ in our conventions (i.e.~with our choice of orientation). Therefore the total time-reversal anomaly of the infrared theory is either $-1$ or $3$, which is in exact agreement with the values found above in the ultraviolet gauge theory. (Analogous matching of the multiple possible values in the ultraviolet of $\nu$ can be carried out in the examples appearing in \cite{Cordova:2017kue,Gomis:2017ixy}.) Indeed, the ultraviolet theory has an unscreened Wilson line in the fundamental representation and the properties of the particle defining the worldline are not part of the definition of the ultraviolet theory. 

A closely related point of view~\cite{Bi:2018xvr} is that we can break the one-form symmetry by adding heavy particles that transform faithfully under the center of the gauge group. We need to assign time-reversal transformations to these new particles. The fundamental Wilson line is now screened by these particles. In particular,  $\nu\in \mathbb{Z}_{16}$ is well defined and the transformation properties of the infrared semions are completely determined in the presence of these heavy particles.\footnote{We thank T. Senthil for discussions on this.}

The mathematical interpretation\footnote{We thank R.~Thorngren for collaborating with us on this result.}
 is that one can show that in the presence of a $\mathbb Z_2$ one-form symmetry the value of $\nu$ can be shifted by a change of variables involving the two-form gauge field and $w_1^2$. This can lead to $\Delta \nu=4$ which is precisely what we have found above. Ideally in such theories we should compute the full anomaly polynomial involving the two-form gauge field $B$ and the time-reversal gauge field $w_1$, and this should be compared across renormalization group  flows and dualities. We leave this for the future. 

An additional subtlety that arises in our theories (but not in the theories with an adjoint Majorana) at $m=k=0$ is the presence of a massless scalar field. In the presence of such a gapless mode, it is nontrivial to evaluate the infrared contribution to $\nu$ (and to various other discrete anomalies).

 To avoid the complications of having the Nambu-Goldstone boson when we study the theory on $\mathrm{pin}_+$ manifolds, we could try to add a baryon to the Lagrangian, in such a way that some time-reversal symmetry remains and at the same time the Nambu-Goldstone boson would be lifted due to the explicit breaking of baryon symmetry.

This brings us back to the discussion of which baryon, in fact, condenses (see e.g.~\cite{Bolognesi:2009vm} for a discussion of some baryons in such theories). It turns out that the baryons in these theories are not as simple as one may initially expect. For instance, the naive baryons for the rank-two representation that were discussed in the literature 
\begin{equation}
\epsilon_{i_1\cdots i_N}\epsilon_{j_1\cdots j_N} \psi^{i_1j_1}\cdots \psi^{i_Nj_N}
\end{equation}
vanish identically because of Fermi statistics: 
\begin{equation}
\epsilon_{i_1\dots i_N}\epsilon_{j_1\cdots j_N} \psi^{i_1j_1}\psi^{i_2j_2}\psi^{i_3j_3}\cdots\equiv0
\end{equation}
for arbitrary values of the spinor indices and regardless of which other insertions are used. To construct a baryon with minimal baryon charge one therefore has to add various derivatives, insertions of the field strength or mesons. It would be nice to return to this in the future.

\section*{Acknowledgments}

We would like to thank A.~Armoni, C.~C\'ordova, T.~Johnson-Freyd, P.S.~Hsin, D.~Radi\v cevi\'c, N.~Seiberg, T.~Senthil, and R.~Thorngren for useful discussions. C.C.~and Z.K.~are grateful for the hospitality of the Perimeter Institute for Theoretical Physics at the final stages of this work. The research of D.D.~and J.G.~was supported by the Perimeter Institute for Theoretical Physics. Research at Perimeter Institute is supported by the Government of Canada through Industry Canada and by the Province of Ontario through the Ministry of Economic Development and Innovation. Z.K.~is supported in part by the Simons Foundation grant 488657 (Simons Collaboration on the Non-Perturbative Bootstrap). Any opinions, findings, and conclusions or recommendations expressed in this material are those of the authors and do not necessarily reflect the views of the funding agencies. 

 \vfill\eject

\printbibliography
\end{document}